\documentclass[journal]{IEEEtran}
\usepackage{amsthm}
\usepackage{amsfonts}
\usepackage{amssymb}
\usepackage{makecell}
\ifCLASSINFOpdf
  \usepackage[pdftex]{graphicx}
  \usepackage{graphicx}
  \graphicspath{{../pdf/}{../jpeg/}{../png/}}
  \DeclareGraphicsExtensions{.pdf,.jpeg,.png}
\else
  \usepackage[dvips]{graphicx}
  \graphicspath{{../eps/}}
  \DeclareGraphicsExtensions{.eps}
\fi
\ifCLASSOPTIONcompsoc
  \usepackage[caption=false,font=normalsize,labelfont=sf,textfont=sf]{subfig}
\else
\usepackage[caption=false,font=footnotesize]{subfig}
\fi
\usepackage{multirow}
\usepackage{balance}
\usepackage{multicol}
\usepackage{epstopdf}
\usepackage{verbatim} 

\usepackage[english]{babel}
\usepackage{float}
\usepackage{xcolor}
\usepackage[linesnumbered,ruled,vlined]{algorithm2e}

\usepackage[hidelinks]{hyperref}
\usepackage{cite}
\ifCLASSINFOpdf
   \usepackage[pdftex]{graphicx}
   \else
\usepackage[dvips]{graphicx}
\fi
\usepackage[cmex10]{amsmath}
\usepackage[T1]{fontenc}
\newcommand*{\affmark}[1][*]{\textsuperscript{\dag}}

\begin{document}
\title{Terahertz Beamforming and Group Sparse Channel Estimation Relying on Low-Resolution ADCs in MU Hybrid MIMO systems}

\author{
\normalsize{Abhisha~Garg,~\IEEEmembership{Graduate Student Member,~IEEE,} Suraj~Srivastava,~\IEEEmembership{Member,~IEEE}, Akash~Kumar, Nimish~Yadav, Aditya~K.~Jagannatham,~\IEEEmembership{Senior Member,~IEEE} and Lajos Hanzo,~\IEEEmembership{Life Fellow,~IEEE}}

\thanks{The work is supported by IEEE SPS scholarship grant for $2023, 2024$ and $2025$. The work of Aditya K. Jagannatham was supported in part by the Qualcomm Innovation Fellowship; in part by the Qualcomm 6G UR Gift; in part by the Arun Kumar Chair Professorship. The work of S. Srivastava was supported in part by Anusandhan National Research Foundation’s PM-ECRG/2024/478/ENS; and in part by Telecom Technology Development Fund (TTDF) under Grant TTDF/6G/368. S. Srivastava and A. K. Jagannatham jointly acknowledge the funding support provided by Anusandhan National Research Foundation's Advanced Research Grant ANRF/ARG/2025/005895/ENS. The work of Lajos Hanzo is supported by Engineering and Physical Sciences Research Council (EPSRC) projects is gratefully acknowledged: Platform for Driving Ultimate Connectivity (TITAN) (EP/X04047X/1; EP/Y037243/1); Robust and Reliable Quantum Computing (RoaRQ, EP/W032635/1); PerCom (EP/X012301/1); India-UK Intelligent Spectrum Innovation ICON UKRI-1859. S. Srivastava, A. K. Jagannatham and Lajos Hanzo jointly acknowledge the funding support provided to ICON-project by DST and UKRI-EPSRC under India-UK Joint opportunity in Telecommunications Research.

Abhisha Garg and Aditya K. Jagannatham are with the Department of Electrical Engineering, Indian Institute of Technology Kanpur, Kanpur-208016, India (e-mail: abhisha20@iitk.ac.in; adityaj@iitk.ac.in). 

Suraj Srivastava is with the Department of Electrical Engineering, Indian Institute of Technology Jodhpur, Jodhpur, Rajasthan 342030, India (e-mail: surajsri@iitj.ac.in).

Akash Kumar is with Qualcomm India Pvt. Ltd., Hyderabad, Telangana, 500081, India (email: akkum@qti.qualcomm.com)

Nimish Yadav is with Samsung Semiconductor India Research, Bengaluru, Karnataka, 560048, India (email: nimish.y@samsung.com).

L. Hanzo is with the School of Electronics and Computer Science, University of Southampton, Southampton SO17 1BJ, U.K. (email:lh@ecs.soton.ac.uk)}}
\maketitle
\begin{abstract}
A unified beamforming and channel estimation framework relying on Bayesian learning is conceived. Recognizing the limitations imposed by low-resolution analog-to-digital converter (ADCs) and frequency-dependent propagation effects occurring in the Terahertz (THz) band, we formulate a dual-wideband channel model incorporating root raised cosine (RRC) pulse shaping. To address the non-linear distortions introduced by low-resolution ADCs, Bussgang decomposition is employed, leading to a tractable linearized inference process. By leveraging the shared sparsity inherent in a multi-user (MU) scenario of THz systems, we propose a Hierarchical Bayesian Group-sparse Regression (HBG-SR) based channel learning technique that exploits the \textit{group-sparse} structure of THz band channels. The estimated dominant angle-of-arrival/ angle-of-departure (AoA/AoD) indices are then exploited for appropriately configuring the true-time-delay (TTD) elements in the hybrid transceiver, enabling precise beam alignment across subcarriers and the effective compensation of the beam-squint effect occurring in wideband THz systems. Extensive simulation results validate the efficiency of the proposed channel estimator and the TTD-aided beamforming architecture, highlighting their robustness and performance gains under practical wideband THz system constraints.
\end{abstract}
\vspace{-3mm}
\section{Introduction} \label{intro}
The THz band, spanning from $(0.1-10)$ THz, offers spectral windows characterized by bandwidths of $10$ GHz or higher \cite{jornet2011channel}, thus standing out as a promising avenue of delivering ultra-high data rates and massive short-range connectivity in next-generation wireless communications. In comparison to the mmWave band that spans $30-100$ GHz \cite{venugopal2017channel}, the challenges in this band include having fewer spatial degrees-of-freedom (SDoF) \cite{sarieddeen2021overview}, high probability of blockages and the spatial wideband effect. The short wavelength of the THz signals leads to variation in the delay across different antenna elements within the array, resulting in the deleterious spatial wideband effect \cite{wang2018spatial}. Furthermore, the wideband THz signal exhibits frequency-selective characteristics stemming from the multipath delay spread, consequently inducing a hostile frequency-wideband effect. A general THz system experiencing simultaneous frequency- and spatial-wideband effects is termed a dual wideband system \cite{garg2024angularly}. The so-called \textit{beam-squint} effect arises because of the variation in the effective angle of arrival (AoA)/ angle of departure (AoD) across the subcarriers, which in turn affects the array response vector \cite{wang2018spatial}. This phenomenon leads to a significant difference with respect to its mmWave counterpart and gives rise to a fundamental challenge in this band. In order to avoid this loss, it is critical to incorporate the beam-squint effect in the analog beamformer/ combiner design procedure, which poses a formidable challenge. To address this impediment, this treatise develops novel hybrid transceiver design algorithms for multi-user (MU) THz MIMO systems, while considering the specific scattering characteristics of dual-wideband THz channels \cite{wang2018spatial}. It is also worth noting that accurate knowledge of the channel state information (CSI) is crucial for the design of the transmit precoder/ receive combiner (TPC/ RC) pair and for beam-squint mitigation. Therefore, the development of channel learning techniques for a spatial-wideband THz channels is a worthwhile endeavor, which is addressed in this paper. Table-\ref{acrn} summarizes the acronyms used in this work.

{To fully exploit the capabilities of the THz band, selecting a waveform tailored to THz-specific propagation and hardware constraints is essential. Effective waveform design must jointly consider performance metrics such as peak-to-average power ratio (PAPR), baseband computational complexity, and sensitivity to hardware and RF impairments \cite{sarieddeen2021overview}. In this context, the preference between SC and multi-carrier (MC) waveforms remains actively debated for THz systems. The adoption of SC modulation in the first sub-THz communication standard (IEEE $802.15.3$d) \cite{ieee2017ieee} reinforces its suitability for this regime, as it aligns well with stringent hardware constraints. In particular, single carrier- frequency domain equilization (SC-FDE) architectures provide lower transmitter-side complexity than orthogonal frequency division multiplexing (OFDM) and inherently achieve reduced PAPR, thereby easing the burden on power amplifiers. Furthermore, in the presence of low-resolution ADCs, OFDM-based systems lose subcarrier orthogonality, which significantly degrades OFDM performance and motivates the adoption of SC transmission in this work. Moreover, SC-FDE exhibits improved resilience to THz-specific impairments such as beam misalignment and beam squint \cite{tarboush2022single}, making it a strong candidate for robust and energy-efficient THz communication. Table-V in \cite{tarboush2022single} provides a systematic comparison between SC and MC waveforms and the results demonstrate that SC-FDE achieves a favorable balance between complexity, PAPR performance, and robustness to beam-squint impairments, further motivating its consideration as a practical waveform for THz communications.} The next section will discuss the exiting state-of-the-art techniques.
\vspace{-1mm}
\subsection{Review of existing works} \label{review}
\begin{table} 
    \centering
    \vspace{-4mm}
      \caption{List of Acronyms}
      \vspace{-2mm}
\scalebox{0.9}{{
\begin{tabular}{|l|r|}
\hline
    ADC &  Analog-to-digital converter \\ \hline
    AoA/ AoD & Angle of arrival/ angle of departure \\ \hline
    AQNM & Additive quantization noise model \\ \hline
    AWGN &  Additive white Gaussian noise \\ \hline
    BCRLB & Bayesian Cram{\'e}r Rao lower bound \\ \hline
    BER &  Bit error rate \\ \hline
    CS & Compressive sensing \\ \hline
    CSI & Channel state information \\ \hline
    DPP & Delay phase precoding \\ \hline
    EM & Expectation maximization \\ \hline
    FFT & Fast Fourier transform \\ \hline
    GMM & Gaussian mixture model \\ \hline
    GSOMP & Generalized simultaneous orthogonal matching pursuit \\ \hline
    HBG-SR & Hierarchical Bayesian Group-sparse Regression \\ \hline
    HITRAN & High-resolution transmission molecular absorption \\ \hline
    IFFT & Inverse fast Fourier transform \\ \hline
    IRS & Intelligent reflecting surface \\ \hline
    LASSO & Least absolute shrinkage and selection operator \\ \hline
    MC & Multi carrier \\ \hline
    MFOCUSS & MMV Focal underdetermined system solver \\ \hline
    MMSE & Minimum mean squared error \\ \hline
    MMV-LS & Multiple measure vector least sqaures \\ \hline
    MMV-SBL/ MSBL & Multiple measure vector sparse Bayesian learning \\ \hline
    MP & Message passing \\ \hline
    NLoS & Non-line-of-sight \\ \hline
    NMSE & Normalized mean square error \\ \hline
    NOMA & Non-orthogonal multiple access \\ \hline
    OFDM & Orthogonal frequency division multiplexing \\ \hline 
    PAPR & Peak-to-average power ratio \\ \hline
    QoS & Quality of service \\ \hline
    RAs & Receiving antennas \\ \hline
    RC & Receive combiner \\ \hline
    RRC & Root raised cosine \\ \hline
    SBL & Sparse Bayesian learning \\ \hline
    SC-FDE & Single carrier frequency domain equalization \\ \hline
    SDoF & Spatial degrees-of-freedom\\ \hline
    SE & Spectral efficiency \\ \hline
    SMV & Single measurement vector \\ \hline
    SNR & Signal to noise ratio \\ \hline
    SVD & Singular value decomposition \\ \hline
    TAs & Transmit antennas \\ \hline
    TD & Time delay \\ \hline
    TPC & Transmit precoder \\ \hline
    TTD & True-time delay \\ \hline
    \end{tabular}
    }}\vspace{-2 \baselineskip}
     \label{acrn}
\end{table}
The THz band is characterized by severe propagation losses, including molecular absorption, reflection losses, and path loss, which pose significant challenges to reliable communication. In a seminal contribution, Jornet and Akyildiz \cite{jornet2011channel} developed a comprehensive THz channel model that incorporates both propagation and molecular absorption effects. Their work investigates the capacity of THz systems under various molecular compositions and pulse shapes using power allocation schemes. In parallel, Piesiewicz \textit{et al.} \cite{piesiewicz2007scattering} explored the impact of reflections in indoor THz scenarios by measuring the reflection coefficients of common building materials using Kirchhoff scattering theory. Their findings highlight that scattering significantly influences both signal power and propagation behavior in the THz band. Priebe \textit{et al.,} \cite{priebe2011aoa}, in their pioneering treatise, model a double directional indoor THz channel, incorporating domains for the angle of AoA/AoD and time of arrival (ToA).

In their cutting-edge work, Lin and Li \cite{lin2015adaptive} proposed a \textit{distance-aware} adaptive THz beamforming scheme, where long-distance users are allocated the central part of the spectrum, while short-distance users are assigned sidebands to ensure quality of service (QoS). Their hybrid beamforming technique does not account for the beam-squint effect, which is severe in the THz domain. Wan \textit{et al.,} \cite{wan2021hybrid} proposed a hybrid transceiver design based on the alternating direction method of multipliers (ADMM), which compensates for the beam-squint effect in MIMO-aided OFDM systems. However, their framework is designed for the mmWave/sub-THz regime. Gao \textit{et al.,} \cite{gao2021wideband} proposed hybrid transceiver design approaches based on a virtual sub-array and true-time-delay (TTD) to compensate for the beam-squint effect. The virtual sub-array approach relies on beam broadening, while the TTD-aided approach replaces all the phase shifters with a few TTD lines. Dai \textit{et al.,} \cite{dai2022delay}, in their cutting-edge work, developed a delay phase precoding (DPP)-based beamforming architecture. This architecture, mitigates the beam-squint effect by introducing a small number of time delay (TD) elements along with the PSs, providing frequency-dependent phase shifts to alleviate the array gain loss caused by the beam-squint effect. However, none of the papers mentioned above considers a MU scenario, which is critical in practical systems. Moreover, the aforementioned works do not consider the impact of employing low-resolution ADCs. This choice makes detection more challenging at the receiver, particularly due to the beam-squint effect \cite{wang2018spatial}. Furthermore, their beamformer designs are based on utilizing the true channel frequency response, which is unavailable in practical scenarios.

Omid \textit{et al.,} \cite{omid2023robust} proposed a novel hybrid beamforming design, which incorporates the CSI uncertainty in intelligent reflecting surfaces (IRS) assisted non-orthogonal multiple access (NOMA) systems. Jafri \textit{et al.,} \cite{jafri2022robust} proposed a distinctive hybrid beamformer design in the context of a cooperative scenario for mmWave hybrid MIMO systems. Their work underscores the importance of considering realistic CSI uncertainty in cooperative scenarios, which is crucial in the beamformer design. Although, these works consider CSI uncertainty, they do not actually estimate the channel. But again, for the design of a practical beamformer, precise channel knowledge is crucial. 

Moreover, due to the highly directional nature of propagation attained by large-scale antennas, the THz band results in a sparsely populated multipath channel in the angular domain. The problem of finding sparse solutions from a single measurement vector (SMV) has received extensive attention in prior research. In the family of SMV models, sparse Bayesian learning (SBL) is one of the most popular choices due to the global convergence of its cost function \cite{wipf2004sparse}. Wipf and Rao \cite{wipf2004sparse}, in their seminal work, initially introduced the SBL algorithm, specifically designed for SMV problems and subsequently expanded the scope of the SBL algorithm to multiple measurement vector (MMV) problems offering a more comprehensive solution for dealing with complex data sets. 
Dovelos \textit{et al.,} \cite{dovelos2021channel} proposed a generalized simultaneous orthogonal matching pursuit (GSOMP) based technique for estimating the dual-wideband THz channel by considering a system with a single user (SU) relying on a single antenna. Chou \textit{et al.,} \cite{chou2023compressed} proposed an MMV least squares (MMV-LS-CV)-based CSI learning technique for estimating the dual-wideband THz channel in hybrid MIMO-OFDM systems. Although their pioneering work incorporates time selectivity, the LS based solution advocated therein yields a poor performance in comparison to SBL. Kim and Choi \cite{kim2021spatial} proposed a Newtonized FCFGS-CV-based channel estimator for wideband mmWave massive MIMO systems having hybrid architectures and low-resolution ADCs. To account for the propagation delay across the antenna array, the discrete-time channel is formulated by incorporating the spatial-wideband effect. However, their formulation is based on a time-domain channel model, which leads to significant complexity in wideband scenarios. {Note that, existing works \cite{dovelos2021channel}, \cite{chou2023compressed}, \cite{kim2021spatial} have addressed wideband effects in mmWave/ sub-THz/THz systems from different perspectives; however, their scope remains limited when compared to the present study. Specifically, both \cite{dovelos2021channel} and \cite{chou2023compressed} consider dual-wideband channel effects, accounting for the joint frequency selectivity arising from large bandwidths and large antenna arrays. However, these works primarily focus on single-user transmission scenarios and concentrate on channel estimation or training design. On the other hand, \cite{kim2021spatial} investigates spatial-wideband channel estimation under hybrid architectures with low-resolution ADCs, but is limited to mmWave systems and does not account for dual-wideband effects, MU transmission or transceiver design. In contrast, the present work jointly considers MU MIMO transmission in dual-wideband THz systems, incorporates spatial-wideband effects together with low-resolution ADC constraints, and exploits group sparsity for CSI learning, which is further leveraged for beamspace-aware hybrid transceiver design. When all these impairments are jointly considered, the conventional linear assumption becomes invalid, since these effects interact across dimensions in a nonlinear fashion rather than simply adding up.} 

{\textit{Cui et al.} \cite{cui2025ice} in their cutting-edge work proposed a Bayesian regression-based channel estimation framework for dense antenna systems from an information-theoretic perspective. Due to the extremely high spatial correlation inherent in dense arrays, the channel prior covariance significantly deviates from an identity matrix. Therefore, the authors designed the \textit{observation matrix} by aligning it with the dominant subspace of the posterior kernel, thereby minimizing the estimation error through covariance-aware sensing. Importantly, the estimation itself is performed via the posterior mean, i.e., the MMSE estimator, and the core contribution lies in the covariance-driven design of the observation matrix. In contrast, our proposed framework differs fundamentally in both modeling and algorithmic structure. Firstly, instead of estimating a full covariance kernel, we impose a parameterized Gaussian prior directly on the channel coefficients, where sparsity is learned through hyperparameters. This leads to a hierarchical Bayesian inference framework that explicitly exploits structured sparsity rather than covariance alignment. Moreover, the observation matrix in our work is not designed by matching a posterior covariance subspace; instead, it is dictated by practical hybrid analog architectures by phase-only controllable combiners. Additionally, our framework is applicable to hybrid MU-THz systems and accommodates practical hardware constraints such as low-resolution ADCs and beam squint effects. } Therefore, we develop a novel CSI acquisition technique that exploits the \textit{group sparsity} naturally arising in MU THz systems, followed by the design of a TTD-based hybrid transceiver that accounts for the effects of low-resolution ADCs. Table \ref{tab:lit_rev} boldly contrasts our contributions, which are highlighted in more detail below.
\begin{table*}
    \centering
\caption{\small Contrasting our contributions to the existing literature} \label{tab:lit_rev}
\vspace{-2mm}
\begin{tabular}{|l|c|c|c|c|c|c|c|c|c|c|c|c|c|c|c|}

    \hline

\textbf{Features} & \cite{garg2024angularly} & \cite{lin2015adaptive}&\cite{gao2021wideband} &\cite{dai2022delay} &\cite{dovelos2021channel} &\cite{chou2023compressed} &\cite{kim2021spatial}     &    \cite{sha2021channel}&\cite{venugopal2017channel} &\textbf{This Paper} \\

 \hline

Sub-THz/ THz Band

& \checkmark  &  \checkmark &  \checkmark &  \checkmark &  \checkmark  & \checkmark  &  & \checkmark  &    & \checkmark\\

 \hline

Reflection \& molecular absorption losses

&  \checkmark & \checkmark &   &   & \checkmark   & \checkmark  &  &   &    & \checkmark\\

 \hline

Diffused-ray modeling

&  \checkmark  &   &   &   &    &   &  &  &    & \checkmark\\
 \hline

SC-FDE/ SC-FDMA system

&  \checkmark &  &   &   &    &   &  &   &  \checkmark  & \checkmark\\
\hline

AoA/AoD with GMM

&   & \checkmark &   &   &    &   &  &   &  & \checkmark\\
\hline

BCRLB

& \checkmark  &  &   &   &  \checkmark  &   &    & \checkmark &  & \checkmark\\
 \hline

Low Resolution ADCs with spatial-wideband

& &  & &  &  &  & \checkmark &   &  &  \checkmark\\

 \hline

CSI learning in THz domain

&  \checkmark &   &   &   &  \checkmark  & \checkmark  & \checkmark & \checkmark &    & \checkmark\\
 \hline

Transceiver design in THz domain

&   & \checkmark  & \checkmark  & \checkmark  &  \checkmark  &   &  &  &    & \checkmark\\
 \hline

\textbf{Group sparsity based CSI learning}

&   &  &   &   &    &   &  &   &    & \checkmark\\

 \hline

\textbf{MU-MIMO in dual-wideband THz system}

& &  &  &  &  &  &  & &    & \checkmark \\

 \hline

\textbf{RRC-PSF and Rect-PSF based dual-wideband channel}

&    &   &   &   &    &   &  &  &    & \checkmark\\
 \hline

\textbf{Beamspace-aware transceiver design with estimated angular support}

&    &   &   &   &    &   &  &  &    & \checkmark\\
 \hline

\end{tabular} \vspace{-1 \baselineskip}
\end{table*}
\subsection{Contributions} \label{contri}
\begin{enumerate}
    \item We commence by developing a practical MU dual-wideband THz channel model encompassing both frequency and distance-dependent characteristics including free-space, absorption and reflection losses. Furthermore, it integrates the effect of multipath delays and transmit pulse shaping, which is absent in \cite{dovelos2021channel}, \cite{chou2023compressed}. Additionally, we incorporate the modeling of diffused rays, which are present in the THz band, yet they are not considered in the existing literature \cite{sha2021channel}. {Moreover, the AoA/AoDs are generated using Gaussian Mixture Model (GMM) \cite{lin2015adaptive}, and a practical THz angle modeling technique that provides an accurate representation of the THz channel.}
    \item We propose a novel Hierarchical Bayesian Group-sparse Regression (HBG-SR)-based channel learning framework, where the BS jointly processes the signals received from all users and the channel exhibits a shared sparse structure. Owing to the shared spatial geometry across subcarriers, the channel inherently possesses a group-sparse structure that can be exploited for reliable CSI recovery. Accordingly, the proposed HBG-SR technique effectively leverages this group sparsity in the concatenated beamspace channel.
    \item To benchmark the performance of the proposed CSI learning technique, we furthermore derive the Bayesian Cram{\'e}r-Rao lower bound (BCRLB) for the MU THz scenario. This serves as an excellent benchmark for CSI estimation performance analysis.
    \item Moreover, the effect of low-resolution ADCs is incorporated into the MU-THz system to account for practical quantization constraints. Additionally, in the THz band, the beam-squint effect induces frequency-dependent signal variations across the antenna array. When combined with coarse quantization, this poses substantial challenges for reliable signal reconstruction and decoding. We address this challenge in this work.
    \item While TTD-based architectures have been widely studied for beam-squint mitigation, existing studies largely focus on SU scenarios \cite{dai2022delay}, \cite{gao2021wideband}, \cite{dovelos2021channel}. However, in MU systems, the shared analog TTD network must simultaneously serve users roaming in different directions, leading to different delay requirements which results in \textit{unavoidable residual beam squint.} Moreover, in wideband THz systems using low-resolution ADCs, this residual frequency-dependent array gain directly translates into non-uniform quantization distortion, since the quantization noise power scales with the ADC input variance. Therefore, this work addresses the joint delay compensation for all the users by conditioning the wideband signal prior to quantization to mitigate the joint impact of residual beam squint and coarse quantization.
    \item Our simulation results quantify the enhanced performance of the proposed CSI learning and the TTD based hybrid transceiver design both in terms of normalized mean square error (NMSE), bit error rate (BER), and spectral efficiency (SE).
    \end{enumerate}
\begin{figure}
\centering
\includegraphics[scale=0.16]{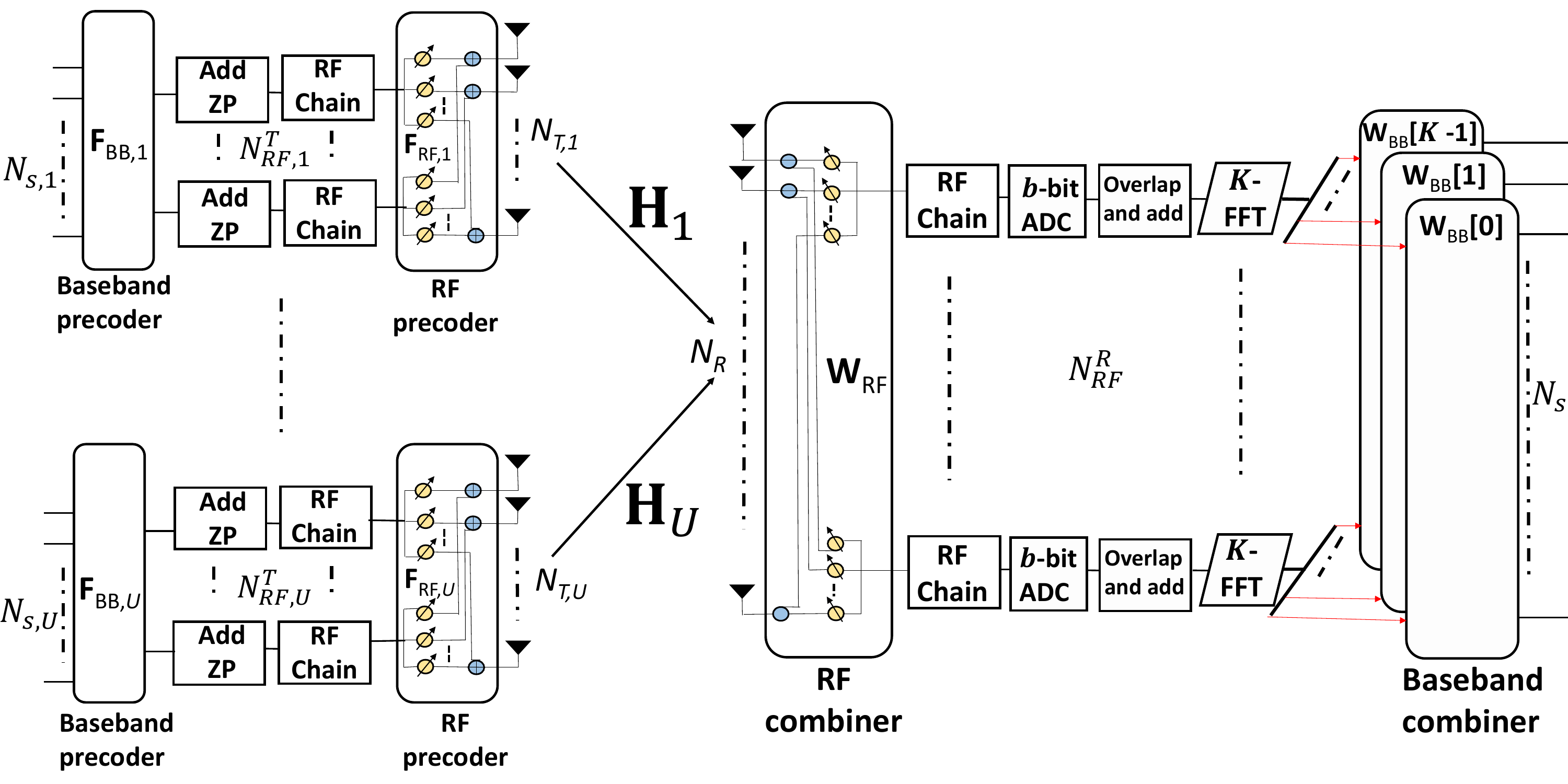}
\vspace{-2mm}
\caption{Schematic diagram of SC-FDE based MU THz hybrid MIMO systems with low-resolution ADCs} \vspace{-1 \baselineskip}
\label{THz_MIMO}
\end{figure}
\subsection{Notations} \label{nota}
Upper-case letters $\mathbf{A}$ represent a matrix, while lower-case letters $\mathbf{a}$ represent a vector. The quantity $\lVert \cdot \rVert_\mathcal{F}$ represents the Frobenius norm, while $\otimes$ represents the Kronecker product and $\mathrm{SVD}(\cdot)$ represent singular value decomposition. The operation $\mathcal{C}(\cdot)$ denotes the column space, and $\mathcal{R}(\cdot)$ denotes the row space of a matrix. The identity $\mathrm{vec}(\mathbf{ABC}) = (\mathbf{C}^T \otimes \mathbf{A})\mathrm{vec}(\mathbf{B})$ illustrates the $\mathrm{vec}(\cdot)$ operation, and $\mathcal{CN}(\boldsymbol{\mu}, \mathbf{\Sigma})$ represents a complex Gaussian distribution with mean $\boldsymbol{\mu}$ and covariance matrix $\mathbf{\Sigma}$. The Dirichlet function is defined as
$\Xi_N(x) = \frac{\sin (\frac{N \pi}{2}x)}{\sin (\frac{\pi}{2}x)}$.
\section{SC-FDE based MU THz Hybrid MIMO System Model} \label{SU_model}
\begin{figure}
\centering
\subfloat[]{\includegraphics[scale=0.5]{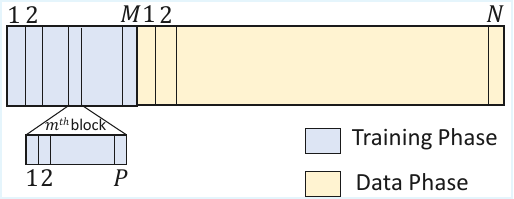}}
\hfil
\hspace{-5pt} \subfloat[]{\includegraphics[scale=0.3]{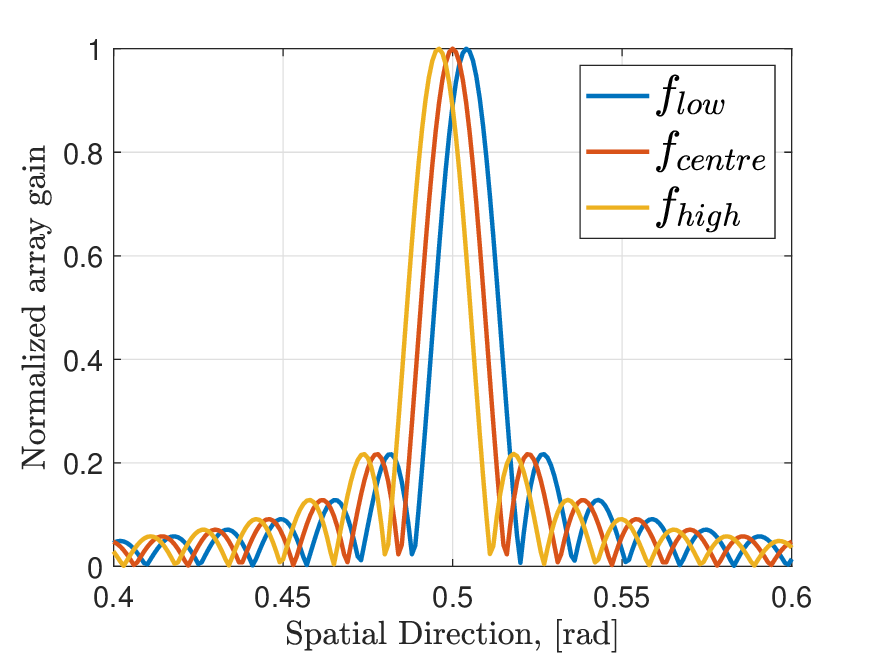}}
\caption{Frame structure utilized for wideband MU THz hybrid MIMO system using SC-FDE} \vspace{-1.2 \baselineskip}
\label{NAG}
\end{figure}

Consider the uplink of a SC wideband THz hybrid MIMO system, where a BS is equipped with $N_{BS}$ receiving antennas (RAs), serving $U$ users simultaneously. Each user is equipped with $N_u$ transmit antennas (TAs), and the antennas of both the BS and user equipment (UEs) are arranged in a uniform linear array (ULA). Let all the users possess equal number of antennas and therefore the total number of TAs is $N_T = \sum_{u=1}^U N_u = UN_u$. Let the THz MIMO system facilitate the transmission of $N_s$ data streams, where each user possesses $N_{s,u}$ data streams. Furthermore, in our consideration of a practical scenario, the BS is equipped with $N_{RF}^B$ radio frequency (RF) chains, while each user is equipped with $N_{RF}^u$ RF chains and follows the relationship of $N_s \leq \sum_{u=1}^U N_{RF}^u \leq N_{RF}^B \ll N_{BS}$ and $N_{RF}^u \ll N_u$. At the UEs, the wideband THz MIMO system comprises a frequency-flat baseband TPC $\mathbf{F}_{\mathrm{BB},u} \in \mathbb{C}^{N_{RF}^u \times N_{s,u}}$ coupled with the frequency-flat RF TPC $\mathbf{F}_{\mathrm{RF},u} \in \mathbb{C}^{N_u \times N_{RF}^u}$ for each user. Furthermore, the BS has a bank of frequency-selective hybrid combiners $\mathbf{W}_{\mathrm{BB}}[k] \in \mathbb{C}^{N_{RF}^B \times N_s}, \: \forall \: 0 \leq k \leq K-1$, where $K$ represents the number of subcarriers. This is cascaded with the frequency-flat RF RC $\mathbf{W}_{\mathrm{RF}} \in \mathbb{C}^{N_{BS} \times N_{RF}^B}$ as shown in Fig. \ref{THz_MIMO}. Moreover, the elements of the RF TPC and RC matrices obey a unit-magnitude constraint and are set as $\left|\mathbf{F}_{\mathrm{RF},u}(\ell,\kappa)\right| = \frac{1}{\sqrt{N_u}}$, $\left|\mathbf{W}_{\mathrm{RF}}(\ell,\kappa)\right| = \frac{1}{\sqrt{N_{BS}}} \: \forall \: \ell,\kappa$. Let us further consider a wideband THz channel consisting of $D$ delay taps at each user, where the $d$-th delay tap can be expressed by a complex matrix $\mathbf{H}_{d,u} \in \mathbb{C}^{N_{BS} \times N_u}, \forall \: 0 \leq d \leq D-1$. As depicted in Fig. \ref{NAG}(a), each transmit frame is divided into two phases, which includes the training and data phase. The training frame consists of $M$ blocks, where each block comprises $P$ pilot vectors. Let $\mathbf{b}_{m,u}^{(p)} \in \mathbb{C}^{N_{RF}^u \times 1}$ denote the $p$-th complex pilot vector, where the relationship, $0 \leq p \leq P-1,$ follows for each $m$-th block and $u$-th user. Before transmitting the $\mathbf{b}_{m,u}^{(p)}$ pilot sequences on an individual RF chain, $D-1$ zeros are appended to each block, resulting in a zero-padded (ZP) block of length $K = P+D-1$, which is given by $\big\{\mathbf{g}_{m,u}^{(q)}\big\}_{q=0}^{K-1} = \Big\{\mathbf{g}_{m,u}^{(0)}, \cdots, \mathbf{g}_{m,u}^{(P - 1)},\underbrace{\mathbf{0},\cdots,\mathbf{0}}_{D-1}\Big\}$ \cite{venugopal2017channel}. In a similar vein, the THz MIMO channel $\mathbf{H}_{d,u}$ at each user $u$ is appended with a zero-matrix of size $N_{BS} \times N_u$ to have length $K$, which is given as $\left\{\mathbf{H}_{q,u}\right\}_{q=0}^{K-1} = \Big\{\mathbf{H}_{0,u},\cdots,\mathbf{H}_{D-1,u},\underbrace{\mathbf{0},\cdots,\mathbf{0}}_{P-1}\Big\}$. It is worth noting that although the overall system model is decoupled into different frequency bins at the receiver, interestingly, at the transmitter, it can be viewed as a single carrier system. This is due to the fact that there is no inverse FFT (IFFT) block at the transmitter and the precoding operation is applied uniformly across the entire bandwidth, as a result, the digital precoders at the user side are naturally frequency independent, consistent with the single-carrier transmission structure. Therefore, the received signal $\mathbf{r}_m(q) \in \mathbb{C}^{N_{BS} \times 1}$ corresponding to all the $U$ users is formulated as
\begin{align}
    \mathbf{r}_m(q) = \sum_{u=1}^U \mathbf{H}_{q,u} \otimes_K (\mathbf{F}_{\mathrm{RF},m,u}\mathbf{F}_{\mathrm{BB},m,u}\mathbf{g}_{m,u}^{(q)})+{\mathbf{v}}_m(q), \notag
\end{align}
where ${\mathbf{v}}_m(q)$ represents the complex additive white Gaussian noise (AWGN) following the distribution $\mathcal{CN}(\mathbf{0}_{N_{BS}\times1},\sigma_n^2 \mathbf{I}_{N_{BS}})$. The quantity $\otimes_K$ represents the circular convolution of length $K$, is defined in Section \ref{nota}. Furthermore, the received signal $\mathbf{r}_m(q)$ is passed through the RF RC followed by a $b$-bit quantizer denoted as $\mathcal{Q}(\cdot)$ which is given by
\vspace{-2mm}
\begin{align}
    \tilde{\mathbf{y}}_m(q) = \mathcal{Q}\Big(\mathbf{W}_{\mathrm{RF},m}^H \sum_{u=1}^U\mathbf{H}_{q,u} \otimes_K &(\mathbf{F}_{\mathrm{RF},m,u}\mathbf{F}_{\mathrm{BB},m,u}\mathbf{g}_{m,u}^{(q)})+ \notag \\ &\mathbf{W}_{\mathrm{RF},m}^H{\mathbf{v}}_m(q)\Big). \label{received_signal}
\end{align}
A widely adopted technique for modeling the quantization effect in the wireless system involves the use of a mid-rise quantizer \cite{mo2017channel}. However, in THz systems operating with high bandwidth and a large number of antennas, the mid-rise quantizer becomes computationally intensive and does not scale efficiently \cite{sarieddeen2021overview}. To address this, we employ the Bussgang decomposition \cite{demir2020bussgang}, which offers a tractable and analytically convenient approximation of the quantization nonlinearity. The received signal $\tilde{\mathbf{y}}_m(q) \in \mathbb{C}^{N_{RF}^B \times 1}$ after applying the Bussgang decomposition can be approximated as
\vspace{-2mm}
\begin{align}
    \tilde{\mathbf{y}}_m(q) \approx \mathbf{Q}\mathbf{W}_{\mathrm{RF},m} \sum_{u=1}^U \mathbf{H}_{q,u} \otimes_K \tilde{\mathbf{g}}_{m,u}^{(q)}+\mathbf{Q} \mathbf{W}_{\mathrm{RF},m}^H{\mathbf{v}}_m(q) + \tilde{\mathbf{v}},
\end{align}
where $\tilde{\mathbf{g}}_{m,u}^{(q)}=\mathbf{F}_{\mathrm{RF},m,u}\mathbf{F}_{\mathrm{BB},m,u}\mathbf{g}_{m,u}^{(q)}$ represents the equivalent transmit vector. The quantization matrix $\mathbf{Q} \in \mathbb{C}^{N_{RF}^B \times N_{RF}^B}$ is diagonal in nature \cite{jacobsson2017throughput}, assuming that each RF chain is quantized independently. Moreover, the matrix $\mathbf{Q}$ is expressed as $\mathbf{Q} = \varkappa \mathbf{I}_{N_{RF}^B},$ where $\varkappa = 1-\rho$ and $\rho$ denotes the inverse of the signal to noise power ratio (SNR). Table-\ref{bin} presents the values of $\rho$ corresponding to each quantization bit for $b \leq 5$. For $b>5$, $\rho$ is approximated by \cite{fan2015uplink} $\frac{\pi \sqrt{3}}{2}2^{-2b}$.

Let $\mathbb{E}\{\mathbf{g}_{m,u}^{(p)}(\mathbf{g}_{m,u}^{(p)})^H\} = \sigma_d^2\mathbf{I}_{N_{s,u}}$, where $\sigma_d^2$ represents the pilot power. The quantized noise vector $\tilde{\mathbf{v}}$ follows the distribution $\mathcal{CN}(\mathbf{0}_{N_{RF}^B \times 1},\mathbf{J}_m),$ where $\mathbf{J}_m \in \mathbb{C}^{N_{RF}^B \times N_{RF}^B}$ represents the quantized noise covariance matrix. Its expression is detailed in Appendix-\ref{noise_covariance_derivation}. For notational simplicity, let $\check{\boldsymbol{\eta}}_m(q)= \mathbf{Q}\mathbf{W}_{\mathrm{RF},m}^H\check{\mathbf{v}}_m(q)+\tilde{\mathbf{v}} \in \mathbb{C}^{N_{RF}^B \times 1}$ represent the equivalent noise. Therefore, the equivalent noise covariance $\mathbf{P}_{\eta \eta} = \mathbb{E}\{\check{\boldsymbol{\eta}}_m(q)\check{\boldsymbol{\eta}}_m^H(q) \} \in \mathbb{C}^{N_{RF}^B \times N_{RF}^B}$ can be formulated as $\mathbf{P}_{\eta\eta} = \varkappa^2 \sigma_n^2\mathbf{W}_{\mathrm{RF},m}^H\mathbf{W}_{\mathrm{RF},m}+\mathbf{J}_m$. Note that in wideband MU, hybrid MIMO systems employing independent scalar ADCs, the Bussgang gain reduces to a per-RF chain attenuation and is therefore modeled as diagonal, consistent with prior works such as \cite{wang2018wideband}, \cite{ma2025hybrid}. This follows from the fact that, under the adopted scalar ADC Bussgang approximation, the gain is determined by the marginal second order statistics of each ADC input, whereas the effects of MU interference, wideband propagation, and analog combining are reflected in the pre ADC signal statistics rather than in the gain term. After applying the $K$-point fast Fourier transform (FFT) to the received signal $\{\tilde{\mathbf{y}}_m[k]\}_{k=0}^{K-1} = \mathrm{FFT}(\{\tilde{\mathbf{y}}_m(q)\}_{q=0}^{K-1})$, one obtains
\begin{align}
    \tilde{\mathbf{y}}_m[k] \approx \mathbf{Q}\mathbf{W}_{\mathrm{RF},m}^H\sum_{u=1}^U\mathbf{H}_u[k]\tilde{\mathbf{g}}_{m,u}[k]+\check{\boldsymbol{\eta}}_m[k],
\end{align}
where $\{\tilde{\mathbf{g}}_{m,u}[k]\}_{k=0}^{K-1} = \mathrm{FFT}(\{\tilde{\mathbf{g}}_{m,u}^{(q)}\}_{q=0}^{K-1})$ represents the FFT of the equivalent transmit vector, while $\{\check{\boldsymbol{\eta}}_m[k]\}_{k=0}^{K-1} = \mathrm{FFT}(\{\check{\boldsymbol{\eta}}_m(q)\}_{q=0}^{K-1})$ represents the FFT of equivalent noise. Finally, the received signal ${\mathbf{y}}_m[k] \in \mathbb{C}^{N_s \times 1}$ after baseband combining $\mathbf{W}_{\mathrm{BB},m}[k] \in \mathbb{C}^{N_{RF}^B \times N_s}$ is given by
\begin{align}
    {\mathbf{y}}_m[k] \approx \mathbf{W}_{\mathrm{BB},m}^H[k]&\mathbf{Q}\mathbf{W}_{\mathrm{RF},m}^H\mathbf{H}_\mathcal{U}[k]\check{\mathbf{F}}_{\mathrm{RF},m}\check{\mathbf{F}}_{\mathrm{BB},m}\check{\mathbf{g}}_m[k]+ \notag \\ & \mathbf{W}_{\mathrm{BB},m}^H[k]\check{\boldsymbol{\eta}}_m[k],
\end{align}
where $\mathbf{H}_\mathcal{U}[k] = [\mathbf{H}_1[k] \, \mathbf{H}_2[k] \, \cdots \, \mathbf{H}_U[k]] \in \mathbb{C}^{N_{BS} \times N_T}$ represents the concatenated channel for all users.
\begin{table}[]
    \centering
    \caption{$\rho$ for different ADC bits $b$ \cite{fan2015uplink}}
    \vspace{-2mm}
    \label{bin}
    \begin{tabular}{|c|c|c|c|c|c|}
        \hline
        $b$ &  $1$ & $2$ & $3$ & $4$ &$5$\\ \hline
        $\rho$ & $0.3634$ & $0.1175$ & $0.03454$ & $0.009497$ & $0.002499$\\ \hline
    \end{tabular} \vspace{-1.5 \baselineskip}
\end{table}
The quantities $\check{\mathbf{F}}_{\mathrm{RF},m} \in \mathbb{C}^{N_T \times N_{RF}^U}$ and $\check{\mathbf{F}}_{\mathrm{BB},m} \in \mathbb{C}^{N_{RF}^U \times N_s}$ represent the overall RF TPC and baseband TPC, respectively, for all the users and are given as $\check{\mathbf{F}}_{\mathrm{RF},m}=\mathrm{blkdiag}(\mathbf{F}_{\mathrm{RF},m,1},\cdots,\mathbf{F}_{\mathrm{RF},m,U})$ and $\check{\mathbf{F}}_{\mathrm{BB},m} = \mathrm{blkdiag}(\mathbf{F}_{\mathrm{BB},m,1},\cdots,\mathbf{F}_{\mathrm{BB},m,U})$. Moreover, the stacked pilot vector is $\check{\mathbf{g}}_m[k] = \big[\mathbf{g}_{m,1}^T[k], \, \mathbf{g}_{m,2}^T[k], \cdots, \mathbf{g}_{m,U}^T[k]\big] \in \mathbb{C}^{N_s \times 1}$ \cite{gonzalez2018channel}. After applying the $\mathrm{vec}(\cdot)$ operator as described in Section-\ref{nota}, the above equation can be re-written as
\begin{align}
    \mathbf{y}_m[k] \approx (\check{\mathbf{g}}_m^T[k]\check{\mathbf{F}}_{\mathrm{BB},m}^T & \check{\mathbf{F}}_{\mathrm{RF},m}^T) \otimes (\mathbf{W}_{\mathrm{BB},m}^H[k]\mathbf{Q}\mathbf{W}_{\mathrm{RF},m}^H) \notag \\ & \mathrm{vec}(\mathbf{H}_\mathcal{U}[k])+\mathbf{W}_{\mathrm{BB},m}^H[k]\check{\boldsymbol{\eta}}_m[k], \label{received_output}
\end{align}
where $\mathrm{vec}(\mathbf{H}_\mathcal{U}[k])=\mathbf{h}_{\mathcal{U}}[k] \in \mathbb{C}^{N_{BS}N_T \times 1}$ represents the concatenated vectorized channel, while $\check{\boldsymbol{\Psi}}_{\mathcal{U},m}[k]  = (\check{\mathbf{g}}_m^T[k]\check{\mathbf{F}}_{\mathrm{BB},m}^T  \check{\mathbf{F}}_{\mathrm{RF},m}^T) \otimes (\mathbf{W}_{\mathrm{BB},m}^H[k]\mathbf{Q}\mathbf{W}_{\mathrm{RF},m}^H) \in \mathbb{C}^{N_s \times N_{BS} N_T}$ represents the concatenated sensing matrix for all users. Let the combined effective noise be $\boldsymbol{\eta}_m[k] = \mathbf{W}_{\mathrm{BB},m}^H[k]\check{\boldsymbol{\eta}}_m[k] \in \mathbb{C}^{N_s \times 1}$, which follows the distribution $\mathcal{CN}(\mathbf{0}_{N_s \times 1},\tilde{\mathbf{P}}_{\eta\eta})$ with $\tilde{\mathbf{P}}_{\eta\eta}=\mathbf{W}_{\mathrm{BB},m}^H[k]\mathbf{P}_{\eta\eta}\mathbf{W}_{\mathrm{BB},m}^H[k]$. Then, the sum spectral efficiency can be expressed as
\begin{align}
    \mathrm{SE} = \frac{1}{K} \sum_{k=1}^K \mathrm{log}_2 \big|\mathbf{I}_{N_s} + \frac{1}{N_s} \tilde{\mathbf{P}}_{\eta\eta}^{-1}\mathbf{H}_{\mathrm{eff},m}[k]\mathbf{H}_{\mathrm{eff},m}^H[k] \big|,
\end{align}
where $\mathbf{H}_{\mathrm{eff},m} = \mathbf{W}_m^H[k]\mathbf{H}_\mathcal{U}[k]\mathbf{F}_m$ and $\mathbf{W}_m[k] = \mathbf{W}_{\mathrm{RF},m}[k]\mathbf{Q}\mathbf{W}_{\mathrm{BB},m}$ while $\mathbf{F}_m = \check{\mathbf{F}}_{\mathrm{RF},m}\check{\mathbf{F}}_{\mathrm{BB},m}$. The next section will describe the formulation of dual-wideband THz MIMO channel model. 
\section{MU THz dual-wideband channel model}
In general, a frequency-dependent array response vector is formulated as \cite{venugopal2017channel}
\begin{align}
    \tilde{\mathbf{a}}(\theta,f) = \frac{1}{\sqrt{N}}\big[1, e^{-j\frac{2\pi}{\lambda}d_c \sin\theta}, \cdots, e^{-j\frac{2 \pi}{\lambda}d_c(N-1) \sin{\theta}} \big]^T, \label{normal_array_response}
\end{align}
where $\lambda = \frac{c}{f}$ represents the wavelength, $d_c$ denotes the inter-antenna spacing typically set as $d_c = \frac{c}{2f_c}$, while $f_c$ represents the carrier frequency. The expression \eqref{normal_array_response} holds for mmWave/sub-THz scenarios where the bandwidth obeys $B \ll f_c$, it does not hold for a wideband THz system. The normalized array gain, as defined in \cite{dovelos2021channel}, is given by
\begin{align}
    \Upsilon = \left|\tilde{\mathbf{a}}^H(\theta_k, f_k) \tilde{\mathbf{a}}(\theta, f_c)\right|^2, \label{nor_arr_gain}
\end{align}
which is plotted in Fig. \ref{NAG}(b) for such a wideband THz MIMO system, where $\mathit{f}_{\mathit{k}} \triangleq \mathit{f}_{\mathit{c}} + \left(\mathit{k} - \frac{\mathit{K} + 1}{2}\right)\frac{\mathit{B}}{\mathit{K}}$ represents the frequency of the $k$-th subcarrier. Observe from Eq. \eqref{nor_arr_gain}, that $\tilde{\mathbf{a}}(\theta_k,f_k)\neq\tilde{\mathbf{a}}(\theta,f_c)$ results in $\Upsilon \neq 1$, which in turn causes the beam to point in different physical directions, giving rise to \textit{beam-squint effects}. Therefore, the modified array response vector \cite{wang2018spatial} after incorporating the beam-squint effect can be expressed as
\vspace{-2mm}
\begin{align}
    \tilde{\mathbf{a}}(\theta,f_k) = \frac{1}{\sqrt{N}}  \big[1, e^{-j\pi \varrho_k \sin{\theta}}, \cdots, e^{-j\pi \varrho_k(N-1) \sin{\theta}} \big]^T, \label{array_res}
\end{align}
where $\varrho_k = \frac{f_k}{f_c}$ represents its relative frequency. Moreover, the maximum array gain is achieved only when $\varrho_k \theta_{k} - \theta = 0$. The derivation of this is detailed in Appendix-\ref{MAG}. The inherent beam-squint effect of wideband THz MIMO systems, causes frequency-dependent beam misalignment and array gain degradation as discussed above. In Section-\ref{Dpp} we will propose an optimal beamformer design. Explicitly, this will replace the conventional frequency-flat analog beamformer by a frequency-dependent one. The next subsection discusses a practical transmit pulse-shaping filter (PSF)-based dual-wideband channel formulation.
\begin{table}
    \centering
    \caption{Notation and description of channel parameters considered}
    \vspace{-2mm}
    \label{notatn}
    \resizebox{0.5\textwidth}{!}{
    {\begin{tabular}{|c|l|c|r|} 
        \hline
        \textbf{Parameter} &  \textbf{Description} & \textbf{Parameter} & \textbf{Description}\\ \hline
        $d$ &  delay tap & $\sigma_n^2$ & noise power\\ \hline
        $K$ &  \# of FFT points & $b$ & quantizer bit\\ \hline
        $\sigma_d^2$ & pilot power & $\rho$ &  inverse SNR of quantization bit \\ \hline
        $\lambda$ &  wavelength & $d_c$ & inter-antenna spacing\\ \hline
        $\Upsilon$ &  normalized array gain & $\rho_k$ & relative frequency\\ \hline
        $\alpha$ &  complex-path gain & $\mathrm{pul}(.)$ & pulse-shaping filter \\ \hline
        $T_s$ &  sampling time & $\mu_{\mathrm{abs}}$ & molecular absorption coefficient  \\ \hline
        $\theta_{(.)}$ & angle of arrival & $\gamma$ & Fresnel reflection coefficient \\ \hline
        $N_{\mathrm{ray}}$ & $\#$ of diffused rays & $N_{\mathrm{NLoS}}$ & $\#$ of multipath components \\ \hline
        $L_{\mathrm{spr}}$ & free-space path loss & $L_\mathrm{abs}$ & molecular absorption loss \\ \hline
        $\tau_{(.)}$ & delay & $\Gamma$ & first-order reflection coefficient \\ \hline
        $\phi_{(.)}$ & angle of departure & $\varphi$ & Rayleigh roughness factor \\ \hline
        $Z(f_k)$ & characteristic impedance & $\omega_i$ & angle of incidence \\ \hline
        $\omega_r$ & angle of refraction & $Z_0$ & intrinsic impedance \\ \hline
        $\mu_0$ & free-space permeability & $\sigma_r$ & standard deviation of material roughness \\ \hline
        $\epsilon_0$ & free-space permittivity & $\xi$ & absorption coefficient of the material \\ \hline
        $ds$ & transmission distance & $\varpi$ & phase of complex path gain \\ \hline
    \end{tabular}}  }
    \\
        \tiny{Note that, the notation $(\cdot)$ represents a generalized index placeholder corresponding to specific parameter, depending on the context.} \vspace{-2 \baselineskip}
\end{table}
\vspace{-1mm}
\subsection{PSF-based dual-wideband channel}
As discussed in Section-\ref{intro}, THz signals exhibit highly directional propagation characteristics, but they have both line-of-sight (LoS) and non-line-of-sight (NLoS) components. Consequently, the dual-wideband THz channel experienced by the $k$-th subcarrier corresponding to each user can be expressed as $\mathbf{H}_u[k] = \mathbf{H}_u^{\mathrm{LoS}}[k] + \mathbf{H}_u^{\mathrm{NLoS}}[k]$, where we have
\begin{align}
\mathbf{H}_u^{\mathrm{LoS}}[k] = {\sqrt{{N_u N_{BS}}}}\alpha(f_k,ds)\beta_{\tau} \mathfrak{B}_{T,u} \mathfrak{B}_R \tilde{\mathbf{a}}_R(\theta,f_k)\tilde{\mathbf {a}}_T^H(\phi,f_k), \label{HLos_sub}
\end{align}
\vspace{-4mm}
\begin{align}
\mathbf{H}_u^{\mathrm{NLoS}}[k] = &{\sqrt{\frac{N_u N_{BS}}{\mathit{N}_{\mathrm{NLoS}}N_{\mathrm{ray}}}}} \sum_{z=1}^{N_{\mathrm{NLoS}}} \sum_{j=1}^{N_{\mathrm{ray}}} \alpha_{z,j}(f_k,ds_{z,j}) \beta_{\tau_{z,j}} \notag \\ & \mathfrak{B}_{T,u}\mathfrak{B}_R \tilde{\mathbf{a}}_R(\theta_{z,j},f_k)\tilde{\mathbf{a}}_T^{H}(\phi_{z,j},f_k). \label{NLoS_channel}
\end{align}
In \eqref{NLoS_channel} we have $\beta_{\tau_{z,j}} = \sum_{d=0}^{\mathit{K}-1}\mathrm{pul}(dT_s-\tau_{z,j})e^{-j\frac{2\pi k d}K},  \forall \, k,d$, where $\mathrm{pul}(\cdot)$ represents the PSF, while $N_{\mathrm{ray}}$ denotes the number of diffused rays within a cluster and $N_{\mathrm{NLoS}}$ represents the number of multipath components. The receive and transmit antenna gains are represented by $\mathfrak{B}_R$ and $\mathfrak{B}_{T,u},$ while the quantities $\theta_{(.)}$, $\phi_{(.)}$, $\tau_{(.)}$, $\alpha_{(.)}$ represent the AoA, AoD, delay and complex path gain. Note that, $\mathfrak{B}_{T,u}/\mathfrak{B}_r$ are angle-dependent antenna gains \cite{tang2020wireless}, \cite{balanis2015antenna}. However, for analytical tractability and to remain consistent with the signal processing literature \cite{dovelos2021channel, lin2015adaptive}, the explicit angle dependence expression is not shown. Nonetheless, in the simulation results, the gain is considered in 'dBi'. The complex path gain $\alpha_{(.)}(f_k,ds)$ can be characterized by its magnitude and phase components $\alpha_{(.)}(f_k,ds) = |\alpha_{(.)}(f_k,ds)|e^{j\varpi}$. Furthermore, the magnitude $\alpha(f_k,ds)$ of the complex path gain depends on the distance $ds$ and for a LoS component which is given by \cite{dovelos2021channel}
\begin{align}
    \left|\alpha(f_k,ds)\right|^2 = L_{\mathrm{spr}}(f_k,ds)L_{\mathrm{abs}}(f_k,ds),
\end{align}
where $L_{\mathrm{spr}}(f_k,ds) = \big(\frac{c}{4\pi f_k ds}\big)^2$ represents the free-space path loss and $L_{\mathrm{abs}}(f_k,ds) = e^{-\mu_{\mathrm{abs}}(f_k)ds}$ represents the molecular absorption loss. The quantity $\mu_{\mathrm{abs}}$ represents the molecular absorption coefficient, which can be calculated from the \textbf{HITRAN} database \cite{hill}. Similarly, the complex path gain of the NLoS components in the $z$-th multipath at the $j$-th diffused ray is given as
\begin{align}
     \left|\alpha_{z,j}(f_k,ds_{z,j})\right|^2 = \Gamma^2_{z,j}(f_k) L_{\mathrm{spr}}(f_k,ds_{z,j})L_{\mathrm{abs}}(f_k,ds_{z,j}),
\end{align}
where $\Gamma_{z,j}(f_k)$ represents the first-order reflection, which is given by the product of Fresnel reflection coefficient $\gamma_{z,j}(f_k) = \frac{Z(f_k) \cos({\omega}_{i_{z,j}}) - Z_0 (\cos{\omega}_{r_{z,j}})}{Z(f_k) \cos({\omega}_{i_{z,j}}) + Z_0 (\cos{\omega}_{r_{z,j}})}$ and Rayleigh roughness factor $ \varphi_{z,j}(f_k) = e^{-\frac{1}{2}\big(\frac{4 \pi f_k \sigma_{r} \cos{\omega}_{i_{z,j}}}{c}\big)^2}$ \cite{piesiewicz2007scattering}. The quantity $\omega_{i_{z,j}}$ denotes the angle of incidence, while the angle of refraction is given by $\omega_{r_{z,j}} = \sin^{-1}\big(\sin(\omega_{i_{z,j}})\frac{\mathit{Z}(\mathit{f}_{\mathit{k}})}{\mathit{Z}_{0}}\big)$, where $Z_0 = 377 \Omega$ is the intrinsic impedance of the free space, and $\sigma_r$ represents the standard deviation of the material. Note that the reason for explicitly considering these losses at THz frequencies is that different materials exhibit distinct electromagnetic behavior due to their frequency-dependent refractive indices and absorption characteristics. These effects significantly influence wave propagation and are modeled using the material-specific parameters \cite{piesiewicz2007properties}, \cite{piesiewicz2007scattering} provided in Table-\ref{material_properties}. Moreover, the quantity $Z(f_k) = \sqrt{\frac{\mu_{0}}{\varepsilon_{0}(n^{2} - (\frac{\xi c}{4\pi f_k})^2 - j\frac{2n\xi c}{4\pi f_k})}}$ represents the characteristic impedance of the reflecting medium \cite{piesiewicz2007scattering}, where $\xi$ represents the absorption coefficient and $n$ is the refractive index, $\mu_{0}$ represents the free space permeability and $\varepsilon_{0}$ denotes the free space permittivity, respectively. Table-\ref{notatn} lists the notation used throughout the work. At this juncture, it is important to note that a THz MIMO channel typically exhibits a limited number of multipath components. Consequently, the next section explores its extended sparse representation tailored for MU THz hybrid MIMO systems.
\section{MU sparse channel estimation model}
Let $G_{T,u}$ and $G_{BS}$ represent the transmit and receive angular bins, which follow the relationship $G_{T,u} \gg N_u, G_{BS} \gg N_{BS}$. Moreover, let $\Theta_{BS}$ and $\Phi_{T,u}$ represent the set of receive and transmit directional sine values \cite{srivastava2021fast}
\begin{align}
\Theta_{BS} = \big\{\theta_{BS}:\sin(\theta_{BS}) = \frac{\mathrm{2}}{\mathit{G}_{BS}}(r-1) - 1, 1\leq r \leq\mathit{G}_{BS} \big\}, \notag
\end{align}
\vspace{-12pt}
\begin{align}
\Phi_{T,u} = \big\{\phi_{t,u}: \sin(\phi_{t,u})=\frac{2}{G_{T,u}}(t-1)-1,1 \leq t \leq G_{T,u} \big\}. \notag 
\end{align}
Therefore, the extended virtual channel model \cite{venugopal2017channel} corresponding to the $u$-th user can be expressed as $\mathbf{H}_u[k] = \mathbf{A}_{BS}(\Theta_{BS},f_k) \mathbf{H}_{b,u}[k] \mathbf{A}_{T,u}^H(\Phi_{T,u},f_k)$. The quantities $\mathbf{A}_{BS}(\Theta_{BS},f_k) \in \mathbb{C}^{N_{BS} \times G_{BS}}$ and $\mathbf{A}_{T,u}(\Phi_{T,u},f_k) \in \mathbb{C}^{N_u \times G_{T,u}}$ represent the receive and transmit array manifold dictionary matrices corresponding to the $u$-th user, which are further given by
\begin{align}
    \mathbf{A}_B(\Theta_{BS}, f_k) = [\tilde{\mathbf{a}}_R(\theta_1, f_k), \tilde{\mathbf{a}}_R(\theta_2, f_k), \cdots, \tilde{\mathbf{a}}_R(\theta_{G_{BS}}, f_k)], \notag
\end{align}
\vspace{-12pt}
\begin{align}
    \mathbf{A}_{T,u}(\Phi_{T,u}, f_k) = \big[\tilde{\mathbf{a}}_{T,u}(\phi_{1,u}, f_k),  \cdots, \tilde{\mathbf{a}}_{T,u}(\theta_{G_{T,u},u}, f_k)\big], \notag
\end{align}
while $\mathbf{H}_{b,u}[k] \in \mathbb{C}^{G_{BS} \times G_{T,u}}$ represents the beamspace domain sparse channel. Leveraging the property of $\mathrm{vec}(\cdot)$ operator as defined in Section-\ref{nota}, the extended virtual channel can be re-written as
\begin{align}
    \mathrm{vec}(\mathbf{H}_u[k]) = \underbrace{[\mathbf{A}_{T,u}^{*}(\Phi_{T,u},f_k) \otimes \mathbf{A}_B(\Theta_{BS}, f_k)]}_{{\tilde{\boldsymbol{\Delta}}}_u[k]} \mathrm{vec}(\mathbf{H}_{b,u}[k]), \notag
\end{align}
where ${\tilde{\boldsymbol{\Delta}}}_u[k] \in \mathbb{C}^{N_{BS}N_u \times G_{BS}G_{T,u}}$ represents the \textit{sparsifying-dictionary}. Furthermore, to formulate a MU dual-wideband vectorized THz hybrid MIMO channel $\mathbf{h}_u[k]$, we concatenate the sparsifying dictionary ${\tilde{\boldsymbol{\Delta}}}_u[k]$ corresponding to each user in block-diagonal fashion, which is mathematically expressed as
\vspace{-2mm}
\begin{align}
    \mathbf{h}_{\mathcal{U}}[k] = \text{blkdiag}\underbrace{\big(\tilde{\boldsymbol{\Delta}}_1[k],\tilde{\boldsymbol{\Delta}}_2[k],\cdots,\tilde{\boldsymbol{\Delta}}_U[k]\big)}_{{\boldsymbol{\Delta}}_{\mathcal{U}}[k]} \underbrace{\begin{bmatrix}
        \mathbf{h}_{b,1}[k] \\ \mathbf{h}_{b,2}[k] \\ \vdots \\ \mathbf{h}_{b,U}[k]
    \end{bmatrix}}_{\mathbf{h}_{b,\mathcal{U}}[k]}. \label{h_con}
\end{align}
The quantity $\boldsymbol{\Delta}_{\mathcal{U}}[k] \in \mathbb{C}^{N_{BS}N_T \times G_{BS}G_T}$ represents the concatenated sparsifying dictionary across all the users, where we introduce $\sum_{u=1}^UG_{T,u} = G_T$ for notational simplicity while the quantity $\mathbf{h}_{b,\mathcal{U}}[k] \in \mathbb{C}^{G_{BS}G_{T} \times 1}$ in \eqref{h_con} represents the concatenated beamspace output. Revisiting \eqref{received_output} again, the combined received output for the $m$-th block at the $k$-th subcarrier across all the users can be mathematically derived as
\vspace{-2mm}
\begin{align}
    \mathbf{y}_m[k] \approx \boldsymbol{\Omega}_{\mathcal{U},m}[k]\mathbf{h}_{b,\mathcal{U}}[k] + \boldsymbol{\eta}_m[k], \label{m_final}
\end{align}
where $\boldsymbol{\Omega}_{\mathcal{U},m}[k]  = \check{\boldsymbol{\Psi}}_{\mathcal{U},m}[k] \boldsymbol{\Delta}_\mathcal{U}[k] \in \mathbb{C}^{N_s \times G_{BS} G_{T}}$ represents the equivalent sensing matrix. Therefore, the MU THz channel estimation model after concatenating the received pilot outputs across all the $M$ blocks is given by
\begin{align}
    \underbrace{\begin{bmatrix}
        \mathbf{y}_{1}[k] \\ \mathbf{y}_{2}[k] \\ \vdots \\ \mathbf{y}_{M}[k]
    \end{bmatrix}}_{\mathbf{y}_{\mathcal{U}}[k]} = \underbrace{\begin{bmatrix}
        \mathbf{\Omega}_{\mathcal{U},1}[k] \\ \mathbf{\Omega}_{\mathcal{U},2}[k] \\ \vdots \\ \mathbf{\Omega}_{\mathcal{U},M}[k]
    \end{bmatrix}}_{\mathbf{\Omega}_{\mathcal{U}}[k]} \mathbf{h}_{b,\mathcal{U}}[k] + \underbrace{\begin{bmatrix}
        \boldsymbol{\eta}_{1}[k] \\ \boldsymbol{\eta}_{2}[k] \\ \vdots \\ \boldsymbol{\eta}_{M}[k]
    \end{bmatrix}}_{\boldsymbol{\eta}_{\mathcal{U}}[k]}, \label{final_system}
\end{align}
where $\mathbf{y}_{\mathcal{U}}[k] \in \mathbb{C}^{MN_s \times 1}$ denotes the stacked received output. Furthermore, $\mathbf{\Omega}_{\mathcal{U}}[k] \in \mathbb{C}^{MN_s \times G_{BS}G_T}$ represents the stacked equivalent sensing matrix, while $\boldsymbol{\eta}_{\mathcal{U}}[k] \in \mathbb{C}^{MN_s \times 1}$ denotes the stacked equivalent noise. Moreover, the stacked noise covariance is $\mathbf{R} = \mathbb{E}\{\boldsymbol{\eta}_\mathcal{U}[k]\boldsymbol{\eta}_\mathcal{U}^H[k]\}=\mathrm{blkdiag}(\mathbf{P}_{\eta\eta,1},\mathbf{P}_{\eta\eta,2},\cdots,\mathbf{P}_{\eta\eta,M}) \in \mathbb{C}^{MN_s \times MN_s}$. 

Interestingly, in wideband THz hybrid MIMO systems, the channel vectors across different subcarriers under the frequency-dependent sparsifying dictionary $\boldsymbol{\Omega}_\mathcal{U}[k]$ of Eq. \eqref{final_system} share a common spatial scattering structure, thereby exhibiting \textit{simultaneous sparsity} that can be further exploited for enhancing the channel estimation accuracy. In this regard, we formulate a MMV model by concatenating the pilot output vectors across all $K$ subcarriers, represented as $\mathbf{Y} = [\mathbf{y}_{\mathcal{U}}[1] \: \mathbf{y}_{\mathcal{U}}[2] \: \cdots \: \mathbf{y}_{\mathcal{U}}[K]] \in \mathbb{C}^{MN_s \times K}$. Moreover, the MU concatenated beamspace channel is denoted as $\mathbf{\Lambda}_b =[\mathbf{{h}}_{b,\mathcal{U}}[1] \: \mathbf{{h}}_{b,\mathcal{U}}[2] \: \cdots \: \mathbf{{h}}_{b,\mathcal{U}}[K] ] \in \mathbb{C}^{G_{BS}G_T \times K}$. One can readily observe from the simultaneously sparse formulation that the dominant spatial scattering paths share a common support across all the subcarriers, which naturally motivates recasting the problem into a group-sparse recovery framework for efficient joint channel estimation over the entire frequency band. Therefore, the next section will discuss the framework of group sparse Bayesian regression.
\section{Hierarchical Bayesian Group Sparse Regression based channel estimation} \label{GSBL}
In this section, we introduce a novel CSI learning technique based on hierarchical Bayesian group-sparse regression, specifically designed for MU uplink THz hybrid MIMO systems, where the BS jointly processes the signal received from all the users and the MU CSI exhibits a shared sparse structure. The proposed HBG-SR technique exploits the \textit{group sparse} structure of the concatenated beamspace channel $\boldsymbol{\Lambda}_b$, whose vectorized form is given by $\mathbf{h}_b = \mathrm{vec}(\mathbf{\Lambda}_b^T)$, and exhibits a structured group-wise representation. Let the $g$-th group in $\mathbf{h}_b$ be denoted as $\mathbf{h}_b^g \in \mathbb{C}^{K \times 1}, \: \forall \: 1 \leq g \leq G_{BS} G_T$, which can be further expressed as $\mathbf{h}_b^g = \mathbf{\Lambda}_b(g,:)$. Let the beamspace channel matrix corresponding to each user $u$, concatenated over all subcarriers $K$, be defined as $\mathbf{H}_{b,u} = \left[\mathbf{h}_{b,u}[0] \: \mathbf{h}_{b,u}[1] \: \cdots \: \mathbf{h}_{b,u}[K-1]\right] \in \mathbb{C}^{ G_{BS} G_{T,u} \times K}$. The prior corresponding to the overall beamspace matrix $\mathbf{\Lambda}_b$ is assumed to be
\vspace{-2mm}
\begin{align}
    f(\mathbf{\Lambda}_b;\mathbf{\Gamma}_{\mathcal{U}}) = \prod_{u=1}^U \prod_{g=1}^{G_{BS}G_u}f(\mathbf{H}_{b,u}(g,:);{\gamma}_{g,u}), \label{G-prior}
\end{align}
where the prior distribution $f(\mathbf{H}_{b,u}(g,:);{\gamma}_{g,u})$ associated with the $g$-th group of user $u$ is given by

\vspace{-5mm}
\small
\begin{align}
    f\big(\mathbf{H}_{b,u}(g,:); {\gamma}_{g,u}\big) = \frac{1}{(\pi {\gamma}_{g,u})^K}\mathrm{exp}\Big(-\frac{\left(\mathbf{H}_{b,u}(g,:)\right)^H\mathbf{H}_{b,u}(g,:)}{{\gamma}_{g,u}}\Big). \notag
\end{align}
\normalsize
Still referring to \eqref{G-prior}, ${\gamma}_{g,u}$ denotes the hyperparameter corresponding to the $g$-th group of the $u$-th user, which is assumed to be unknown. Let the hyperparameter matrix $\mathbf{\Gamma}_{\mathcal{U}}$ be defined as $\mathbf{\Gamma}_{\mathcal{U}} = \mathrm{blkdiag}\left(\mathbf{\Gamma}_1 \, \mathbf{\Gamma}_2 \, \cdots \, \mathbf{\Gamma}_U \right) \in \mathbb{R}^{G_{BS}G_T \times G_{BS}G_T},$ where $\mathbf{\Gamma}_u = \mathrm{diag}\left({\gamma}_{1,u} \, {\gamma}_{2,u} \, \cdots \, {\gamma}_{G_{BS} G_u} \right) \in \mathbb{R}^{G_{BS} G_u \times G_{BS} G_u}$, and $\mathbf{H}_{b,u}(g,:) \in \mathbb{C}^{K \times 1}$ represents the $g$-th group corresponding to the $u$-th user. Note that, if the well-known Bernoulli-Gaussian prior \cite{liu2017sparse} is applied within the Bayesian learning framework, the inference becomes intractable and requires additional approximations \cite{rajoriya2023joint} (as derived in Appendix A). Moreover, adopting a Laplacian prior instead of a Gaussian prior fundamentally alters the inference framework, as the resulting posterior does not admit closed-form expressions \cite{park2008bayesian}. Therefore, we assume a Gaussian prior, because the likelihood and prior are conjugates of each other. 

Let the vector of hyperparameters corresponding to the concatenated beamspace channel $\mathbf{\Lambda}_b$ be given by $[\tilde{\gamma}_1 \; \tilde{\gamma}_2 \; \cdots \; \tilde{\gamma}_{G_{BS} G_T}] = \mathrm{diag}(\mathbf{\Gamma}_\mathcal{U})$. Using this, the prior distribution corresponding to the complete beamspace channel matrix $\mathbf{\Lambda}_b$, can be rewritten from Eq. \eqref{G-prior} as
\begin{align}
    f(\mathbf{\Lambda}_b; \, \mathbf{\Gamma}_{\mathcal{U}}) = \prod_{g=1}^{G_{BS} G_T} \frac{1}{(\pi \tilde{\gamma}_g)^K} \mathrm{exp}\Big(-\frac{\mathbf{\Lambda}_b^H(g,:)\mathbf{\Lambda}_b(g,:)}{\tilde{\gamma}_g}\Big) \label{cprior},
\end{align}
where $\tilde{\gamma}_g$ denotes the hyperparameter associated with the $g$-th group. Therefore, to maximize the Bayesian evidence $\mathrm{log} \, f(\mathbf{Y};\mathbf{\Gamma}_\mathcal{U})$, we employ the expectation-maximization (EM) algorithm, which guarantees local convergence and it is used for estimating the hyperparameter matrix $\mathbf{\Gamma}_\mathcal{U}$.

Let the hyperparameter estimate at the $(j-1)$-st EM iteration be denoted by $\widehat{\mathbf{\Gamma}}_{\mathcal{U}}^{(j-1)}$. The $\mathrm{E}$-step of the $j$-th EM iteration computes the expected value of the log-likelihood function $\mathcal{L}(\mathbf{\Gamma}_{\mathcal{U}}|\widehat{\mathbf{\Gamma}}_{\mathcal{U}}^{(j-1)})$ for the complete dataset $({\mathbf{y}}_{\mathcal{U}},{\mathbf{h}}_{b})$, where we have ${\mathbf{y}}_{\mathcal{U}} = \mathrm{vec}\left(\mathbf{Y}^T\right),$ yielding

\vspace{-4mm}
\small
\begin{equation}
\begin{aligned}
    &\mathcal{L}(\mathbf{\Gamma}_\mathcal{U}|\widehat{\mathbf{\Gamma}}_\mathcal{U}^{(j-1)}) = \mathbb{E}_{\mathbf{h}_b|\mathbf{y}_\mathcal{U};\widehat{\mathbf{\Gamma}}_\mathcal{U}^{(j-1)}}\{\mathrm{log} \, f(\mathbf{y}_{\mathcal{U}},\mathbf{h}_b;\mathbf{\Gamma}_\mathcal{U})\}, \\ & = \mathbb{E}_{\mathbf{h}_b|\mathbf{y}_\mathcal{U};\widehat{\mathbf{\Gamma}}_\mathcal{U}^{(j-1)}}\{\mathrm{log} \,[f(\mathbf{y}_\mathcal{U}|\mathbf{h}_b)]+ \mathbb{E}_{\mathbf{h}_b|\mathbf{y}_\mathcal{U};\widehat{\mathbf{\Gamma}}_\mathcal{U}^{(j-1)}} \{\mathrm{log} \, f(\mathbf{h}_b;\mathbf{\Gamma}_\mathcal{U})\}\}. \label{log_lik_prior}
\end{aligned}
\end{equation}
\normalsize
By applying Bayes' rule to Eq. \eqref{log_lik_prior}, the term $\mathbb{E}_{\mathbf{h}_b|\mathbf{y}_\mathcal{U};\widehat{\mathbf{\Gamma}}_{\mathcal{U}}^{(j-1)}}\{\mathrm{log} \, f(\mathbf{y}_\mathcal{U}|\mathbf{h}_b)\}$ becomes independent of the hyperparameter matrix $\mathbf{\Gamma}_{\mathcal{U}}$, and can thus be omitted in the subsequent $\mathrm{M}$-step. As a result, the equivalent optimization problem that maximizes $\mathcal{L}(\mathbf{\Gamma}_{\mathcal{U}}|\widehat{\mathbf{\Gamma}}_{\mathcal{U}}^{(j-1)})$ with respect to $\mathbf{\Gamma}_{\mathcal{U}}$ can be formulated as
\vspace{-1mm}
\begin{align}
    \widehat{\mathbf{\Gamma}}_{\mathcal{U}}^{(j)} = \mathop{\mathrm{arg \, max}}\limits_{\mathbf{\Gamma}_{\mathcal{U}}} \, \mathbb{E}_{\mathbf{h}_b|\mathbf{y}_\mathcal{U};\widehat{\mathbf{\Gamma}}_\mathcal{U}^{(j-1)}}\{\mathrm{log} \, f(\mathbf{h}_b; \mathbf{\Gamma}_{\mathcal{U}}\}.
\end{align}
From Eq. \eqref{G-prior}-\eqref{cprior}, it readily follows that the M-step update for the $g$-th hyperparameter in the $j$-th iteration is given by
\vspace{-1mm}
\begin{align}
    {\widehat{\gamma}}_{g,u}^{(j)} & = \mathop{\mathrm{arg \, max}}\limits_{\gamma_{g,u}} \mathbb{E}_{\mathbf{h}_b|\mathbf{y}_\mathcal{U};\widehat{\mathbf{\Gamma}}_\mathcal{U}^{(j-1)}}\{\mathrm{log} \, f(\mathbf{H}_{b,u}(g,:); {\gamma}_{g,u})\}, \notag \\ & = \mathop{\mathrm{arg \, max}}\limits_{\gamma_{g,u}} \mathbb{E}_{\mathbf{h}_b|\mathbf{y}_\mathcal{U};\widehat{\mathbf{\Gamma}}_\mathcal{U}^{(j-1)}}\Big\{-K \, \mathrm{log}(\gamma_{g,u})- \notag \\ & \quad\quad\quad\quad\frac{1}{\gamma_{g,u}}\mathrm{Tr}\big(\mathbf{H}_{b,u}(g,:)\mathbf{H}_{b,u}^H(g,:)\big)\Big\}
    . \label{M-step}
\end{align}
Upon differentiating the above equation with respect to $\gamma_{g,u}$ and setting the result equal to zero, we obtain
\begin{align}
    \widehat{\gamma}_{g,u} = \frac{1}{K}\mathrm{Tr}\Big(\mathbb{E}\big(\mathbf{H}_{b,u}(g,:)\mathbf{H}^H_{b,u}(g,:)\big)\Big). \label{mtep}
\end{align}
Moreover, the \textit{a posteriori} pdf is $f(\mathbf{\Lambda}_b|\mathbf{y}_\mathcal{U};\widehat{\mathbf{\Gamma}}^{(j-1)}) = \mathcal{CN}\big({\mathbf{H}}_{b,\mathcal{U}}^{(j)},\mathbf{\Sigma}^{(j)}\big),$ where $\widehat{\mathbf{H}}_{b,\mathcal{U}}^{(j)} \in \mathbb{C}^{G_{BS} G_T \times K}$ represents the \textit{a posteriori} mean and $\tilde{\mathbf{\Sigma}}_k^{(j)} \in \mathbb{C}^{G_{BS} G_T  \times G_{BS} G_T }$ denotes the \textit{a posteriori} covariance matrix, which are given by
\vspace{-1mm}
\begin{align}
\tilde{\boldsymbol{\Sigma}}^{(j)}_k  &= \big[(\widehat{\mathbf{\Gamma}}_{\mathcal{U}}^{(j-1)})^{-1}+\mathbf{\Omega}_{\mathcal{U}}^H[k]\mathbf{R}^{-1}\mathbf{\Omega}_{\mathcal{U}}[k]\big]^{-1}, \notag \\
    \widehat{\mathbf{H}}_{b,\mathcal{U}}^{(j)}(:,k) &= \tilde{\boldsymbol{\Sigma}}_k^{(j)}\mathbf{\Omega}_{\mathcal{U}}^H[k]\mathbf{R}^{-1}\mathbf{Y}(:,k).
\end{align}
Moreover, the \textit{a posteriori} mean $\widehat{\mathbf{h}}_{g,u} \in \mathbb{C}^{K \times 1}$ and the \textit{a posteriori} covariance $\mathbf{\Sigma}_{g,u} \in \mathbb{C}^{K \times K}$ corresponding to the $g$-th group of each user are given as
\vspace{-2mm}
\begin{align}
    \mathbf{h}_{g,u}^{(j)} & = [\widehat{\mathbf{H}}_{b,\mathcal{U}}^{(j)}(i_{g,u},1), \cdots, \widehat{\mathbf{H}}_{b,\mathcal{U}}^{(j)}(i_{g,u},K)], \notag \\
    \tilde{\boldsymbol{\Sigma}}_{g,u}^{(j)} & = \mathrm{diag}\big(\tilde{\mathbf{\Sigma}}_k^{(j)}(i_{g,u},i_{g,u}; 1), \cdots, \tilde{\mathbf{\Sigma}}_k^{(j)}(i_{g,u},i_{g,u}; K)\big), \label{user_aposteriori}
\end{align}
\begin{algorithm}[t]
\DontPrintSemicolon 
\KwIn{Pilot output $\mathbf{Y}$, Equivalent sensing matrix $\mathbf{\Xi}$,  noise covariance $\mathbf{R}$, total subcarriers $K$, $\epsilon$ and $\mathit{K}_{\mathrm{max}}$}

\textbf{Initialization:} $\widehat{\mathbf{\Gamma}}_\mathcal{U}^{(0)}= \mathbf{I}_{G_{BS} G_T}$, $\widehat{\mathbf{\Gamma}}^{(-1)}_\mathcal{U} = \mathbf{0}_{G_{BS}G_T}$ and counter $\mathit{j} = \mathrm{1}$ 
 
\While{$\big(\parallel \widehat{\mathbf{\Gamma}}^{(j)}_\mathcal{U} - \widehat{\mathbf{\Gamma}}^{(j-1)}_\mathcal{U} \parallel_{\mathit{F}}^2 \ > \epsilon \ \&\& \ j < \mathit{K}_{\mathrm{max}}\big)$}
{
 
\textbf{E-Step:} Compute the \textit{a posteriori} covariance and mean as

\For{$k = 1,\cdots, K$}
 {
$\tilde{\boldsymbol{\Sigma}}_k^{(j)}=\big[\big(\widehat{\boldsymbol{\Gamma}}_\mathcal{U}^{(j-1)}\big)^{-1} + \mathbf{\Omega}_{\mathcal{U}}^H[k]\mathbf{R}^{-1}\mathbf{\Omega}_{\mathcal{U}}[k] \big]^{-1}$
\begin{equation}
    \begin{aligned}
        \widehat{\mathbf{H}}_{b,\mathcal{U}}^{(j)}(:,k) = \tilde{\mathbf{\Sigma}}_k^{(j)}\mathbf{\Omega}_{\mathcal{U}}^H[k]\mathbf{R}^{-1} \mathbf{Y}(:,k) \notag
    \end{aligned}
\end{equation}
}
 
 \textbf{M-Step:} Update hyperparameters estimates using
 
$\widehat{\gamma}_g^{(j)} = \frac{1}{K} \sum_{k=1}^K\big(\tilde{\mathbf{\Sigma}}^{(j)}(i_g,i_g;k)+|\widehat{\mathbf{H}}_{b,\mathcal{U}}^{(j)}(i_g,k)|^2\big)$

$j \leftarrow j+1$ 
 
}

\textbf{return:~~}{$\widehat{\mathbf{H}}_{\mathit{b},\mathcal{U}}$}
\caption{HBG-SR based MU THz channel estimation}
\label{HBG-SR}
\end{algorithm}
where $i_{g,u} = \tilde{S}_{u-1} + g, \: \forall \: \tilde{S}_{u-1} < g \leq \tilde{S}_u$ represents the global index, and $S_u = \sum_{m=1}^u G_{BS}G_m$. Upon substituting Eq. \eqref{user_aposteriori} into \eqref{mtep}, we obtain the M-step as
\begin{align}
    \widehat{\gamma}_{g,u}^{(j)} = \frac{1}{K}\mathrm{Tr}\Big[\tilde{\boldsymbol{\Sigma}}_{g,u}^{(j)}+\mathbf{h}_{g,u}^{(j)}\big(\mathbf{h}_{g,u}^{(j)}\big)^H\Big],
\end{align}
which can be re-written as
\begin{align}
    \widehat{\gamma}_g^{(j)} = \frac{1}{K} \sum_{k=1}^K\Big(\tilde{\mathbf{\Sigma}}^{(j)}(i_g,i_g;k)+|\widehat{\mathbf{H}}_{b,\mathcal{U}}^{(j)}(i_g,k)|^2\Big),
\end{align}
where $i_g = \tilde{S}_{u(g)-1} +g$ and $\tilde{S}_u$ remain the same as defined earlier. Algorithm-\ref{HBG-SR} summarizes the proposed sparse estimation technique. The procedure is repeated until convergence. Let the concatenated MU transmit dictionary $\mathbf{A}_T(\Phi_T, f_{0:K-1}) \in \mathbb{C}^{N_T \times G_T \times K}$ be formulated as
\vspace{-3mm}
\begin{align}
    \mathbf{A}_T(\Phi_T, f_{0:K-1}) = \mathrm{blkdiag}[&\mathbf{A}_{T,1}(\Phi_{T,1}, f_{0:K-1}), \cdots, \notag \\ & \mathbf{A}_{T,U}(\Phi_{T,u},f_{0:K-1})].
\end{align}
Therefore, the estimated MU THz MIMO channel using the HBG-SR algorithm is given as
\vspace{-2mm}
\begin{align}
    \widehat{\mathbf{H}}_{\mathrm{HBG}} = \mathbf{A}_B(\Theta_{BS},f_{0:K-1}) \mathrm{vec}^{-1}(\widehat{\mathbf{H}}_{b,\mathcal{U}})\mathbf{A}_{T}^H(\Phi_T, f_{0:K-1}). \notag
\end{align}
Note that, conventional MSBL algorithms, which assume a common dictionary matrix across all the subcarriers \cite{srivastava2021data}, do not lead to optimal performance in dual-wideband scenarios. This implies that when we use a common dictionary across all subcarriers, the \textit{sparse locations will vary} for each subcarrier. However, when we employ a dictionary corresponding to each subcarrier, it fixes the sparse locations, resulting in group sparsity in the channel. In essence, the sparse locations remain consistent across different subcarrier dictionaries, despite each subcarrier having its unique dictionary matrix. This characteristic further enhances the robustness of the algorithm, which makes it novel and can be used in practical scenarios.

As detailed in Section-\ref{review}, the design of hybrid beamformers having TTD elements necessitates accurate beam-steering angles for each RF chain for counteracting the frequency-dependent delays that arise in wideband THz MIMO systems. These delays effectively mitigate the beam-squint by leveraging the angular-domain information embedded in the estimated beamspace channel. Specifically, the proposed HBG-SR inherently promotes sparsity in the channel estimates, wherein the significant non-zero coefficients directly correspond to dominant propagation paths, implicitly encoding precise directional information. Consequently, by prioritizing these coefficients and mapping their indices to the corresponding AoA/AoD pairs, the steering angles required can be efficiently extracted without additional pilot overhead. 

Towards this, Algorithm-\ref{optimal_angles} summarizes the procedure for extracting the dominant AoD/AoA pairs directly from the estimated beamspace channel. Specifically, Step-$3$ extracts the estimated hyperparameter matrix for each user, and Step-$4$ transforms this matrix into a vector by selecting its diagonal elements. Step-$5$ then sorts these hyperparameter values in descending order for ranking their angular significance. Subsequently, Step-$6$ maps the dominant indices of $N_{RF}^u$ hyperparameters onto their corresponding AoD bins for configuring the user-side analog TPC, while Step-$7$ maps the dominant $\frac{N_{RF}^B}{U}$ hyperparameters onto AoA bins for configuring the BS-side analog RC. Finally, Step-$8$ aggregates and returns the selected AoA and AoD indices, representing the dominant transmit and receive angular directions for each user. Consequently, the algorithm circumvents the need to store the complete CSI at the receiver or to fully feed it back to the transmitter, thereby reducing complexity and making it well-suited for practical implementations.
\begin{algorithm}[t]
\DontPrintSemicolon 
\KwIn{$\widehat{\mathbf{H}}_{b,\mathcal{U}}, N_{RF}^u, N_{RF}^B$}
\textbf{Initialization:} $\tilde{\Phi}_T = [\quad], \tilde{\Theta}_{BS} = [\quad], E = [\quad]$
 
\For{$u = 1:U$}
{
$\widehat{\mathbf{h}}_u = \widehat{\mathbf{H}}_{b,\mathcal{U}}[:,(u-1)G_{T,u}+1:uG_{T,u}]$

$[\sim,\tilde{\mathrm{I}}] = \mathrm{sort}(|\widehat{\mathbf{h}}_u|,\mathrm{descend})$

$[\sim,\grave{\Phi}_T] = \mathrm{ind2sub}(G_{BS},G_T,\tilde{\mathrm{I}}(1:N_{RF}^u))$

$[\grave{\Theta}_{BS},\sim] = \mathrm{ind2sub}(G_{BS},G_T,\tilde{\mathrm{I}}(1:\frac{N_{RF}^B}{U}))$

$E_u = I(1:N_{RF}^u)$
  
$\tilde{\Phi}_T = [\tilde{\Phi}_T \,\, \grave{\Phi}_{T}]; \tilde{\Theta}_{BS} = [\tilde{\Theta}_{BS} \,\, \grave{\Theta}_{BS}]; E = [E \,\, E_u]$
}
    
\textbf{Output:~~}{$\tilde{\Phi}_T, \tilde{\Theta}_{BS}, E$}
\caption{Determination of optimal physical angles from the estimated beamspace} 
\label{optimal_angles}
\end{algorithm}
\vspace{-3mm}
\subsection{MU Bayesian Cram{\'e}r-Rao lower bound}
Let us consider the parameterized Gaussian prior of the beamspace matrix $\mathbf{\Lambda}_b,$ as defined in Eq. \eqref{cprior}. Upon taking the natural logarithm on both sides of the Eq. \eqref{cprior}, we obtain 
\vspace{-2mm}
\begin{equation}
    \mathrm{log}\{f(\mathbf{\Lambda}_b; \mathbf{\Gamma}_{\mathcal{U}})\} \! = \!\! \sum_{g=1}^{G_{BS}G_T} \!\! \Big(-K\, \mathrm{log}(\pi \tilde{\gamma}_g) - \frac{1}{\tilde{\gamma}_g}\mathbf{\Lambda}_b^H(g,:)\mathbf{\Lambda}_b(g,:)\Big).\label{log_pror}
    \end{equation}
To express the log-likelihood of the complete beamspace matrix $\mathbf{\Lambda}_b$, Eq. \eqref{log_pror} can   be rewritten as
\begin{equation}
    \begin{aligned}
    \mathrm{log}\{f(&\mathbf{\Lambda}_b; \mathbf{\Gamma}_\mathcal{U})\} = -K \sum_{g=1}^{G_{BS}G_T} \log (\tilde{\gamma}_g) - \\ &\sum_{g=1}^{G_{BS}G_T} \frac{\mathbf{\Lambda}_b^H(g,:)\mathbf{\Lambda}_b(g,:)}{\tilde{\gamma}_g} - GK\log (\pi). \label{beamspace_loglikli}
\end{aligned}
\end{equation}
Let $\sum_{g=1}^{G_{BS}G_T} \tilde{\gamma}_g^{-1} \mathbf{\Lambda}_b^H(g,:)\mathbf{\Lambda}(g,:) = \mathrm{Tr}(\mathbf{\Lambda}_b^H\mathbf{\Gamma}^{-1}_\mathcal{U}\mathbf{\Lambda}_b)$. Therefore, Eq. \eqref{beamspace_loglikli} can be re-written as
\begin{align}
    \log f(\mathbf{\Lambda}_b; \mathbf{\Gamma}_\mathcal{U}) \! = \! -K \log \mathrm{det} (\mathbf{\Gamma}_\mathcal{U}) - \mathrm{Tr}(\mathbf{\Lambda}_b^H \mathbf{\Gamma}_\mathcal{U}^{-1} \mathbf{\Lambda}_b) + \mathrm{cons}.
\end{align}
Thus, the conditional PDF $f(\mathbf{Y}|\mathbf{\Lambda}_b)$ corresponding to the received output $\mathbf{Y}$ is given by $\prod_{k=1}^K\mathcal{CN}\left(\mathbf{y}_\mathcal{U}[k]; \mathbf{\Omega}_\mathcal{U}[k] \mathbf{h}_{b,\mathcal{U}}[k], \mathbf{R}\right)$. Moreover, the log-likelihood of the conditional PDF $f(\mathbf{Y}|\mathbf{\Lambda}_b)$ is given by
\begin{align}
     \mathrm{log}\{f\left(\mathbf{Y}|\mathbf{\Lambda}_b\right)\} = -\sum_{k=1}^K&\Big[\log \mathrm{det}(\mathbf{R}) +  \big(\mathbf{y}_\mathcal{U}[k] - \mathbf{\Omega}_\mathcal{U}\mathbf{h}_b\big)^H \times  \notag\\ &\mathbf{R}^{-1}\big(\mathbf{y}_\mathcal{U}[k] - \mathbf{\Omega}_\mathcal{U}\mathbf{h}_b\big)\Big]. \label{output_loglikli}
\end{align}
Let $\mathbf{J} \in \mathbb{C}^{G_{BS} G_T K \times G_{BS} G_T K}$ denote the Bayesian Fisher Information Matrix (FIM) formulated as
\begin{align}
    \mathbf{J} = \mathbf{J}_y + \mathbf{J}_b,
\end{align}
where $\mathbf{J}_y \in \mathbb{C}^{G_{BS} G_T K \times G_{BS} G_T K}$ denotes the FIM associated with the combined received output $\mathbf{Y},$ while $\mathbf{J}_b \in \mathbb{C}^{G_{BS} G_T K \times G_{BS} G_T K}$ represents the FIM corresponding to the \textit{a priori} information of the beamspace CSI $\mathbf{\Lambda}_b$. Thus, the quantities $\mathbf{J}_y$ and $\mathbf{J}_b$ \cite{srivastava2021data} can be defined as
\begin{align}
    \mathbf{J}_y = -\mathbb{E}_{\mathbf{Y},\mathbf{\Lambda}_b}\Big\{\frac{\partial^2\mathcal{L}\left(\mathbf{Y}|\mathbf{\Lambda}_b\right)}{\partial \mathbf{\Lambda}_b \, \partial \mathbf{\Lambda}_b^H }\Big\}, \mathbf{J}_b = -\mathbb{E}_{\mathbf{\Lambda}_b}\Big\{\frac{\partial^2 \mathcal{L}\left(\mathbf{\Lambda}_b;\mathbf{\Gamma}_{\mathrm{MU}}\right)}{\partial\mathbf{\Lambda}_b \, \partial\mathbf{\Lambda}_b^H}\Big\}, \notag
\end{align}
where $\mathcal{L}(\cdot)$ represents the $\mathrm{log}(f(\cdot))$. By solving Eqs. \eqref{beamspace_loglikli} and \eqref{output_loglikli}, one can obtain the corresponding FIM as
\vspace{-2mm}
\begin{align}
    \mathbf{J}_y &= \mathrm{blkdiag}\big(\mathbf{\Omega}_\mathcal{U}^H[1]\mathbf{R}^{-1}\mathbf{\Omega}_\mathcal{U}[1],\cdots,\mathbf{\Omega}_\mathcal{U}^H[K]\mathbf{R}^{-1}\mathbf{\Omega}_\mathcal{U}[K] \big), \notag \\ \mathbf{J}_b &= \widehat{\mathbf{\Gamma}}_\mathcal{U}^{-1} \otimes \mathbf{I}_K. \label{combine_FIM}
\end{align}
Using the expressions for the received output FIM and the beamspace FIM from equation \eqref{combine_FIM}, the Bayesian FIM can therefore be obtained as 
\begin{align}
    \mathbf{J} = \mathrm{blkdiag}\big(\mathbf{\Omega}_\mathcal{U}^H[1]\mathbf{R}^{-1}\mathbf{\Omega}_\mathcal{U}[1],\cdots,&\mathbf{\Omega}_\mathcal{U}^H[K]\mathbf{R}^{-1}\mathbf{\Omega}_\mathcal{U}[K] \big) + \notag \\ &\widehat{\mathbf{\Gamma}}_\mathcal{U}^{-1} \otimes \mathbf{I}_K. \label{bayesian_FIM}
\end{align}
Let the concatenated equivalent sparsifying dictionary be defined as $\mathbf{S} = \mathrm{blkdiag}(\boldsymbol{\Delta}_\mathcal{U}[1] \: \boldsymbol{\Delta}_{\mathcal{U}}[2] \: \cdots \: \boldsymbol{\Delta}_{\mathcal{U}}[K]) \in \mathbb{C}^{N_{BS} N_T K \times G_{BS} G_T K}$. Finally, the BCRLB for the MSE estimate of $\widehat{\mathbf{\Lambda}}_b$ can be formulated as
\begin{align}
    \mathrm{MSE}\big\{\widehat{\mathbf{\Lambda}}_b\big\} \geq \mathrm{Tr}\left\{\mathbf{S}\,\mathbf{J}^{-1}\,\mathbf{S}^H\right\}.
\end{align}
\subsection{Computational complexity}
This subsection derives the computational complexities of the HBG-SR, GSOMP and MMV-LS based THz channel learning techniques. Table-\ref{LS1} list the computational complexity of the MMV-LS scheme. As described in \cite{kay1993statistical}, the MMV-LS scheme possesses a worst-case computational complexity of $\mathcal{O}(N_{BS}^3N_T^3)$ flops, primarily arising from the inversion and multiplication operations involved in the LS estimation process. Table-\ref{GSMplexity} list the per-subcarrier and per-iteration complexity of the GSOMP framework along with various key computational steps. Note that, the GSOMP algorithm follows the same algorithmic structure as done in our earlier work \cite{11250920}. Moreover, it applies to a generic MMV formulation and is not restated explicitly to avoid redundancy. The associated computational complexity is, however, included to enable comparison and complexity performance trade-offs. The overall computational complexity of the GSOMP framework is on the order of $\left(KG_{BS}G_TMN_s+K(\frac{2}{3}j^3+j^2MN_s+3jMN_s)\right)$, where $j$ represents the current iteration. The worst-case complexity order is $\mathcal{O}(KM^3N_s^3),$ which is attributed to the necessity of an intermediate LS estimate at each iteration. Table-\ref{BGlexity} lists the per-subcarrier and per-iteration complexity of the HBG-SR framework along with various key computational steps. The worst-case computational complexity for all the subcarriers is $\mathcal{O}\big(K(G_{BS}^2G_T^2MN_s + M^3N_s^3)\big),$ implying that the computational complexity increases linearly with the number of subcarriers and also higher than GSOMP and MMV-LS framework. However, as shown in Section-VII, HBG-SR indicates superior performance. Hence, there is a trade-off between computational complexity and estimation performance.
\begin{table*}[hbt!]
    \centering
    \caption{Computation complexity of MMV-LS}
    \label{LS1}
    \vspace{-2mm}
    \resizebox{0.7\textwidth}{!}{
    \begin{tabular}{|l|r|r|}
        \hline
        \textbf{Operation} &  \textbf{Complex multiplications} & \textbf{Complex additions} \\ \hline
        $\mathbf{\Psi}_\mathcal{U}^H[k]\mathbf{\Psi}_\mathcal{U}[k]$ & $\frac{MN_s N_{BS}N_T(N_{BS}N_T+1)}{2}$ & $\frac{(MN_s-1)N_{BS}N_T(N_{BS}N_T+1)}{2}$ \\ \hline
        $(\mathbf{\Psi}_\mathcal{U}^H[k]\mathbf{\Psi}_\mathcal{U}[k])^{-1}$ & $\frac{N_{BS}^3N_T^3}{2}+\frac{3N_{BS}^2N_T^2}{2}$ & $\frac{N_{BS}^3N_T^3}{2}-\frac{3N_{BS}^2N_T^2}{2}$ \\ \hline
        $(\mathbf{\Psi}_\mathcal{U}^H[k]\mathbf{\Psi}_\mathcal{U}[k])^{-1}\mathbf{\Psi}_\mathcal{U}^H[k]$ & $N_{BS}^2N_T^2MN_s$ & $N_{BS}N_TMN_s(N_{BS}N_T-1)$ \\ \hline
        $(\mathbf{\Psi}_\mathcal{U}^H[k]\mathbf{\Psi}_\mathcal{U}[k])^{-1}\mathbf{\Psi}_\mathcal{U}^H[k]\mathbf{Y}_{\mathrm{MU}}(:,k)$ & $N_{BS}N_TMN_s$ & $N_{BS}N_T(MN_s-1)$ \\ \hline
    \end{tabular}}\vspace{-1 \baselineskip}
\end{table*}
\begin{table*}[hbt!]
    \centering
    \vspace{-2mm}
    \caption{Computation complexity of GSOMP scheme, per-subcarrier in the $j$-th iteration}
    \vspace{-2mm}
    \label{GSMplexity}
    \resizebox{0.7\textwidth}{!}{
    \begin{tabular}{|l|r|r|}
        \hline
        \textbf{Operation} &  \textbf{Complex multiplications} & \textbf{Complex additions} \\ \hline
        $\mathbf{\Omega}_\mathcal{U}^H[k]\mathbf{T}(:,k)$ & $G_{BS}G_TMN_s$ & $G_{BS}G_T(MN_s-1)$ \\ \hline
        $\widehat{\mathbf{H}}_{\mathrm{LS},\, \mathcal{U}}(:,k) = \left(\mathbf{\Omega}^\mathcal{A}_\mathcal{U}\right)^\dagger\mathbf{Y}_\mathcal{U}(:,k)$ & $\frac{2}{3}j^3 + j^2MN_s+2jMN_s$ & $\frac{2}{3}j^3+j^2(MN_s-1)+j(MN_s-1)$ \\ \hline
        $\mathbf{Y}_\mathcal{U}(:,k)-\mathbf{\Omega}_\mathcal{U}^\mathcal{A}\widehat{\mathbf{H}}_{\mathrm{LS}, \,\mathcal{U}}(:,k)$ & $jMN_s$ & $jMN_s$ \\ \hline
    \end{tabular}}\vspace{-1 \baselineskip}
\end{table*}
\begin{table*}[hbt!]
    \centering
    \caption{Computation complexity of BGSR framework, per-subcarrier, per-EM iteration}
    \vspace{-2mm}
    \label{BGlexity}\resizebox{0.7\textwidth}{!}{
    \begin{tabular}{|l|r|r|}
        \hline
        \textbf{Operation} &  \textbf{Complex multiplication} & \textbf{Complex additions} \\ \hline
        $(\widehat{\mathbf{\Gamma}}_\mathcal{U}^{(j-1)})^{-1}$ & $\frac{G_{BS}^3G_T^3}{2}+\frac{3 G_{BS}^2 G_T^2}{2}$ & $\frac{G_{BS}^3G_T^3}{2}-\frac{3 G_{BS}^2 G_T^2}{2}$\\ \hline
        $\mathbf{R}^{-1}$ & $\frac{M^3N_s^3}{2} + \frac{3M^3N_s^3}{2}$ & $\frac{M^3N_s^3}{2} - \frac{3M^3N_s^3}{2}$ \\ \hline
        $\mathbf{\Omega}_\mathcal{U}^H[k]\mathbf{R}^{-1}$ & $G_{BS}G_TM^2N_s^2$ & $G_{BS}G_TM^2N_s^2-G_{BS}G_TMN_s$\\ \hline
        $\mathbf{\Omega}_\mathcal{U}^H[k]\mathbf{R}^{-1}\mathbf{\Omega}_\mathcal{U}[k]$ & $G_{BS}^2G_T^2MN_s$ & $G_{BS}^2G_T^2MN_s-G_{BS}^2G_T^2$ \\ \hline
        
        $\big[(\widehat{\mathbf{\Gamma}}_\mathcal{U}^{(j-1)})^{-1} +$ $\mathbf{\Omega}_\mathcal{U}^H[k]\mathbf{R}^{-1}\mathbf{\Omega}_\mathcal{U}[k]\big]^{-1}$ & $\frac{G_{BS}^3G_T^3}{2}+\frac{3 G_{BS}^2 G_T^2}{2}$ & $\frac{G_{BS}^3G_T^3}{2}-\frac{3 G_{BS}^2 G_T^2}{2}$ \\ \hline
        $\mathbf{\Omega}_\mathcal{U}^H[k]\mathbf{R}^{-1}\mathbf{Y}(:,k)$ & $G_B G_T M N_s$ & $G_B G_T (M N_s-1)$ \\ \hline
        $\tilde{\boldsymbol{\Sigma}}_k^{(j)}\mathbf{\Omega}_\mathcal{U}^H[k]\mathbf{R}^{-1}\mathbf{Y}(:,k)$ & $G_B^2G_T^2$ & $G_BG_T(G_BG_T-1)$ \\ \hline
    \end{tabular}}\vspace{-1 \baselineskip}
\end{table*}
\vspace{-2mm}
\section{TTD based Hybrid Combiner design}\label{Dpp}
In this section, we develop a beam-squint mitigation strategy in which frequency-dependent TTD elements perform coarse delay compensation across the signal bandwidth, followed by a bank of frequency-flat PSs that refine the residual steering error. Consequently, the overall beamformer design is structured into two stages; the first stage involves identifying the optimal frequency-independent beam steering angles, while the second stage utilizes these angles to construct frequency-dependent RF TPCs and RCs. Furthermore, to determine the angles in the first stage, we leverage the estimated beamspace AoA/AoD pairs obtained from Algorithm-\ref{optimal_angles} to configure the TTD elements. 
\subsection{Generation of frequency dependent beamformers}
Let each RF chain be connected to $S$ TD elements, which are further connected to $P=\frac{N}{S}$ PSs. Upon revisiting Eq. \eqref{normal_array_response} and dividing the array response vector into $S$ equal number of elements, we obtain
\vspace{-2mm}
\begin{align}
    \tilde{\mathbf{a}}_{1,k}^l = \tilde{\mathbf{a}}_1^l e^{-j2\pi f_kt_1^l}, \cdots, \tilde{\mathbf{a}}_{S,k}^l = \tilde{\mathbf{a}}_S^l e^{-j2\pi f_kt_S^l},
\end{align}
where $\tilde{\mathbf{a}}_{(.)}^T \in \mathbb{C}^{P \times 1}$ and $\tilde{\mathbf{a}}(\theta_k,f_k) = [\tilde{\mathbf{a}}_1^T, \cdots, \tilde{\mathbf{a}}_S^T]^T \in \mathbb{C}^{N \times 1}$. Let $\mathbf{t}^l = [t_1^l, \cdots, t_S^l]^T \in \mathbb{C}^{S \times 1}$ represent the delays corresponding to each RF chain. Therefore, the overall frequency-dependent PS corresponding to the $k$-th subcarrier can be expressed as
\vspace{-2mm}
\begin{align}
    \tilde{\mathbf{a}}_k^l   = \mathrm{blkdiag}((\tilde{\mathbf{a}}_1^l)^T, \cdots, (\tilde{\mathbf{a}}_S^l)^T)e^{-j 2 \pi f_k \mathbf{t}^l}. \label{arry_res_TTD}
\end{align}
Let $\mathbf{p}_k^l = e^{-j 2 \pi f_k \mathbf{t}^l}$ represent the frequency-dependent PS. To maintain the directivity, one can generate a frequency-dependent beam to share the same form as the array response vector $\tilde{\mathbf{a}}_k$ expressed as
\begin{align}
    \mathbf{p}_k^l = [1, e^{-j \pi v_k^l},  \cdots, e^{-j \pi (S-1) v_k^l}]^T, \label{direc_vector}
\end{align}
where $v_k^l \in [-1,1]$ represents the directional rotation vector. Therefore, by adjusting $v_k^l$, the beams generated by $\tilde{\mathbf{a}}_k^l$ can be aligned with the target physical direction $\theta^l$ corresponding to all the subcarriers $k$. Moreover, the maximum array gain is achieved at an angle $\omega_{\mathrm{opt}}$, when
\begin{align}
    \omega_{\mathrm{opt}} = \underset{\theta^l}{\mathrm{arg \, max}} |(\tilde{\mathbf{a}}(\theta^l,f_c))^H \tilde{\mathbf{a}}(\theta_k^l,f_k)|,
\end{align}
which further reduces to 
\begin{align}
    \omega_{\mathrm{opt}} = \frac{\theta^l}{\varrho_k} - \frac{v_k^l}{P\varrho_k}, \label{optimal_angle}
\end{align}
as detailed in Appendix-\ref{opti_cal}. Thus, aligning the optimal beam steering angle with the actual physical direction yields $v_k = (1-\varrho_k)\theta P$, implying that the beam-squint effect can be largely compensated using the TD elements. Additionally, it can be observed from Eq. \eqref{direc_vector} that the phase differences between adjacent TTD elements are equal and implying that the TD vector $\mathbf{t}$ should satisfy the following form
\begin{align}
    \mathbf{t}^l = [0, nT_c, \cdots, (S-1)nT_c]^T, \label{time_delay}
\end{align}
where $T_c = \frac{1}{f_c}$ represents the time-period of the carrier frequency and $n$ represents number of periods corresponding to each RF chain.
\vspace{-2mm}
\subsection{Calculation of optimal time delays}
Upon substituting the TD value in the expression of the frequency-dependent PS, one obtains
\begin{align}
    \mathbf{p}_k^l = \big[1, e^{-j 2 \pi f_k n T_c}, \cdots, e^{-j 2 \pi f_k (S-1) n T_c}\big]^T.
\end{align}
When comparing it to Eq. \eqref{direc_vector}, the time period obtained becomes $n = \frac{P\theta^l}{2 \varrho_k}(\varrho_k-1)$. In addition, one can observe that the number of periods depends not only on the fixed PS $P$ and the physical direction $\theta$, but also on the relative frequency $\varrho_k$. Furthermore, substituting the value of $n$ into the TD vector, one arrives at
\begin{align}
    t_s^l = (s-1)nT_c = (s-1)\frac{P\theta^l}{2}T_c - (s-1)\frac{P\theta^l}{2 \varrho_k}T_c,
\end{align}
which represents the first term to be fixed as $\tilde{t}_s^l = (s-1)\frac{P\theta^l}{2}T_c, \, \forall \, 1<s<S,$ for all subcarriers, while the second term can be realized by adding an extra phase shift. Revisiting Eq. \eqref{arry_res_TTD} and substituting the value of TD vector into it, the overall array response vector can be expressed as \eqref{arr_vc}
\begin{figure*}
    \begin{align}
    \tilde{\mathbf{a}}_k^l &= \mathrm{blkdiag}((\tilde{\mathbf{a}}_1^l)^T, (\tilde{\mathbf{a}}_2^l)^T, \cdots, (\tilde{\mathbf{a}}_S^l)^T)e^{-j 2 \pi f_k \mathbf{t}^l}, \notag \\ &= \mathrm{blkdiag}\big((\tilde{\mathbf{a}}_1^l)^T, (\tilde{\mathbf{a}}_2^l)^T e^{j \pi P \theta^l}, \cdots, (\tilde{\mathbf{a}}_S^l)^T e^{j \pi (S-1)P \theta^l}\big) e^{-j2\pi f_k\mathbf{t}^l},  \label{arr_vc}
\end{align}\vspace{-1 \baselineskip}
\hrulefill
\end{figure*}
where the actual time delay obeys $\tilde{\mathbf{t}}^l = [\tilde{t}_1^l, \tilde{t}_2^l, \cdots, \tilde{t}_S^l]$. Therefore, the generalized time delay corresponding to each RF chain can be obtained as
\begin{align}
\tilde{t}_{l,s} =
\begin{cases}
(s-1)\dfrac{P\theta_l}{2}T_c, & \theta_l \ge 0, \\
(S-1)\left|\dfrac{P\theta_l}{2}\right|T_c + (s-1)\dfrac{P\theta_l}{2}T_c, & \theta_l < 0 .
\end{cases}
\label{final_TD}
\end{align}
where $t_{s,l} \in [0,\frac{N}{2}T_c]$ with $\theta_l \in [-1,1]$ and $1 \leq s \leq S$. For a representative configuration with $f_c = 0.65$ THz, $N_{BS} = 64$ and $S = 2$, the corresponding delay range is between $0$ to $50$ ps. The delay values are practically achievable by artificial transmission-line based TTDs and delay-line based TTD architectures \cite{meng2024compact}, \cite{jung2020compact}. These designs support delay ranges and resolutions that are consistent with the values considered in this work. In our analysis, we assume sufficiently fine-grained and practically realizable delay resolution in order to focus on the fundamental signal-processing behavior. At the same time, we do not impose a specific hardware-dependent delay quantization or implementation constraint, as the objective of this work is not to optimize a particular TTD circuit design but to study the system-level interaction between delay compensation and low-resolution ADCs.
Algorithm-\ref{com_biner} summarizes the design procedure of a frequency-dependent RC. A similar procedure can be applied for the frequency-dependent TPC, where the optimal precoder angle $\tilde{\Phi}_T$ is used as input; the detailed procedure is omitted here due to space constraints.
\vspace{-2mm}
\subsection{Minimum number of TD elements required to mitigate the beam-squint effect}
To effectively mitigate the beam-squint effect, it is further essential to determine the minimum number of TD elements required per subarray, ensuring frequency-independent beam alignment across the entire operational bandwidth. Revisiting, the directional rotation vector which includes, $-1 \leq v_k^l \leq 1,$ and substituting the optimal value, we get
\begin{align}
    -1 \leq (1-\varrho_k)P \theta^l \leq 1.
\end{align}
Moreover, to design the worst-case design constraint, we consider $\theta = 1$. Under this condition, the minimum number of TD elements required is given by
\vspace{-2mm}
\begin{align}
    S \geq N(1-\varrho_k),
\end{align}
which represents the theoretical lower bound on $M$. In practical implementations, this bound must be rounded up to the nearest integer to ensure realizability, i.e., the effective number of TD elements is given by $\lceil N(1-\varrho_k) \rceil$.
\begin{algorithm}[t]
\DontPrintSemicolon 
\KwIn{$\widehat{\mathbf{H}}_u[k], \mathbf{A}_{B}(\Theta_{BS},f_c), \tilde{\Theta}_{BS}, E$}
 
\For{$\mathsf{i} = 1, 2, \cdots, N_{RF}^B$}
{

 $\tilde{\mathbf{a}}^{\mathsf{i}} = \mathbf{A}_B(\Theta_{BS},f_c)[:,E(\mathsf{i})]; $
 
 $\theta^{\mathsf{i}} = \Theta_{BS}(:,\tilde{\Theta}_{BS}(\mathsf{i}))$

  \For{$s = 1,2,\ldots,S$}
 {
$\tilde{\mathbf{a}}^{\mathsf{i}}_s = \tilde{\mathbf{a}}^{\mathsf{i}}[1+(s-1)P:sP] e^{j \pi (S-1) P \theta^{\mathsf{i}}}$
 
Calculate TD $\tilde{t}_s^{\mathsf{i}}$ using equation \eqref{final_TD}

$\tilde{\mathbf{a}}_{s,k}^{\mathsf{i}} = \tilde{\mathbf{a}}_s^{\mathsf{i}} e^{-j 2 \pi f_k \tilde{t}_s^{\mathsf{i}}}$
}

$\mathbf{W}_{\mathrm{RF}}[k](:,\mathsf{i}) = \big[(\tilde{\mathbf{a}}_{1,k}^{\mathsf{i}})^T, \cdots,(\tilde{\mathbf{a}}_{S,k}^{\mathsf{i}})^T \big]^T$
 
}

$\tilde{\mathbf{H}}_{\mathrm{eq}}[k] = \mathbf{H}_u^H[k] \mathbf{W}_{\mathrm{RF}}[k]$

$\left[\mathbf{U}_{\mathrm{eq}}[k] \mathbf{\Sigma}_{\mathrm{eq}}[k] \mathbf{V}_{\mathrm{eq}}^H[k]\right] = \mathrm{SVD}(\tilde{\mathbf{H}}_{\mathrm{eq}}[k])$

$\mathbf{W}_{\mathrm{BB},u}[k] = \mathbf{V}_{\mathrm{eq},[:,1:N_s^u]}[k]$

\textbf{return:~~}{$\mathbf{W}_{\mathrm{RF}}[k], \mathbf{W}_{\mathrm{BB},u}[k]$}
\caption{TTD based frequency-dependent combiner design}
\label{com_biner}
\end{algorithm}
\begin{table}
\centering
\caption{Simulation parameters for the hybrid transceiver design and sparse estimation}
\vspace{-2mm}
\label{tab}
\resizebox{0.43\textwidth}{!}{%
\begin{tabular}{|l|c|}\hline
\textbf{Parameter} & \textbf{Value} \\ \hline \hline
$\#$ of TAs at UE ($N_{u}$)  & $4$ \\\hline
$\#$ of RAs ($N_{BS}$)  & $64$ \\ \hline
$\#$ of TAs RF chains ($N_{RF}^u$)  & $2$ \\ \hline
$\#$ of RAs RF chains ($N_{RF}^{B}$)  & $8$ \\ \hline
$\#$ of time delay per RF chain ($S$)  & $2$ \\ \hline
$\#$ of subcarriers ($K$) &  $128$ \\ \hline
$\#$ of pilot blocks ($M$) & $20$ \\ \hline
$\#$ of data vectors ($N_d$)  & $100$ \\ \hline
$\#$ of Users ($U$)  & $3$ \\ \hline
$\#$ of NLoS components/ clusters ($N_{\text{NLoS}}$) &  $3$ \\ \hline
$\#$ of diffused rays ($N_{ray}$) &  $1$ \\ \hline
$\#$ of delay taps ($D$) &  $6$ \\ \hline
Molecular Absorption loss ($K_{\mathrm{abs}}$) &  $0.015$ $\mathrm{m}^{-1}$ \\ \hline
Transmit Antenna Gain at each UE ($\mathfrak{B}_{T,u}$) &  $8 \: \text{dBi}$ \\ \hline
Receive Antenna Gain at each subarray ($\mathfrak{B}_R$) &  $28 \: \text{dBi}$ \\ \hline
Transmit angular grid size at UE ($G_{T,u}$) & $8$ \\ \hline
Receive angular grid size ($G_{BS}$)  & $128$ \\ \hline
Operating frequency ($f_c$) & $0.65 \: \text{THz}$ \\\hline
Transmission distance ($ds$) & $15 \text{m}$ \\\hline
Bandwidth ($B$) & $5 \: \text{GHz}$ \\\hline
Angle quantization parameter ($N_Q$) & $4$ \\\hline
ADC Resolution ($\rho$) & $3$-\text{bit} \\\hline
Constellation & $8$-\text{PSK} \\\hline
Roll-off factor for RRC-PSF & $0.80$ \\\hline
Upsampling factor & $20$ \\ \hline
HBG-SR Threshold ($\varepsilon, N_\mathrm{max}$) & $10^{-4},30$ \\
\hline
\end{tabular}}
\end{table}
\begin{table}
\centering
\caption{List of materials used for the simulation environment \cite{piesiewicz2007scattering},\cite{piesiewicz2007properties}}
\vspace{-2mm}
\label{material_properties}
\resizebox{0.4\textwidth}{!}{%
\begin{tabular}{|c|c|c|c|}\hline
\textbf{Material Type} & $\sigma_r (\mathrm{in \: mm})$ & $\xi (\mathrm{in \: cm^{-1}})$ & $n$ \\\hline
Polycarbonate (PC) & $0$ & $23$ & $1.52$  \\\hline
Polystyrene (PS) & $0.002$ & $2$& $1.6$ \\\hline
Polyvinyl chloride (PVC) & $0.028$ & $19$ & $1.68$  \\\hline
Plaster s1 & $0.05$ & $10$ & $2$ \\\hline
Gypsum plaster & $0.13$ & $38$ & $1.4$  \\\hline
Plaster s2 & $0.15$ & $10$ & $2$  \\\hline
\end{tabular}
} \vspace{-1 \baselineskip}
\end{table}
\section{Simulation Results} \label{sim_on}
This section provides the performance analysis of the proposed TTD based hybrid beamformer and the HBG-SR based channel learning algorithm. Table-\ref{tab} outlines the key numerical parameters used in our simulation study. Note that, for simplicity, we consider only one diffused ray within each cluster. However, the framework can be readily extended to accommodate multiple diffused rays. Additionally, the inter-antenna spacing between the AEs is set to be $\frac{\lambda}{2}$. The number of channel paths is $N_{\mathrm{LoS}}+N_{\mathrm{NLoS}} = 4$. The path positions are assumed to be independent and identically distributed, with their delay $\tau_{(.)}$ chosen uniformly from $[0, (D-1)T_s]$ with $T_s = \frac{1}{B}$ \cite{venugopal2017channel}. We consider an indoor office environment with a molecular composition of $20.9\%$ oxygen, $78.1\%$ nitrogen and $1\%$ water vapor. The material parameters of various scatterers are listed in Table-\ref{material_properties}. Note that in the THz band, the refractive index exhibits frequency-dependent behavior, which causes these materials to interact differently with electromagnetic waves across the frequency spectrum. The SNR in decibels (dB) is calculated as $\mathrm{SNR} = 10 \mathrm{log}_{10}(\frac{1}{\sigma_n^2})$. The transmitting and receiving grid-sizes are considered to be $G_{T,u} \geq 2N_{u}$ and $G_{BS} \geq 2N_{BS}$ \cite{gonzalez2018channel}. The frequency-independent PSs are modeled similar to \cite{garg2024angularly} given as
\vspace{-1.5mm}
\begin{align}
    \mathbf{F}_{\mathrm{RF},m,u}(a,b) = \frac{1}{\sqrt{N_{T,u}}} e^{j\upsilon_{a,b}}, \mathbf{W}_{\mathrm{RF},m} = \frac{1}{\sqrt{N_R}}e^{j\zeta_{a,b}},
\end{align}
where the phases $\upsilon_{a,b}$ and $ \zeta_{a,b}$ are randomly sampled with uniform probability as $\mathcal{F} = \Big\{0,\frac{2\pi}{2^{N_Q}},\cdots,\frac{(2^{N_Q}-1)}{2^{N_Q}} \Big\}.$
Moreover, the AoAs and AoDs are generated using GMM to provide a more accurate spatial representation \cite{lin2015adaptive}, with their second-order statistics \cite{priebe2011aoa} given by
\begin{figure*}
	\centering
\subfloat[]{\includegraphics[scale=0.45]{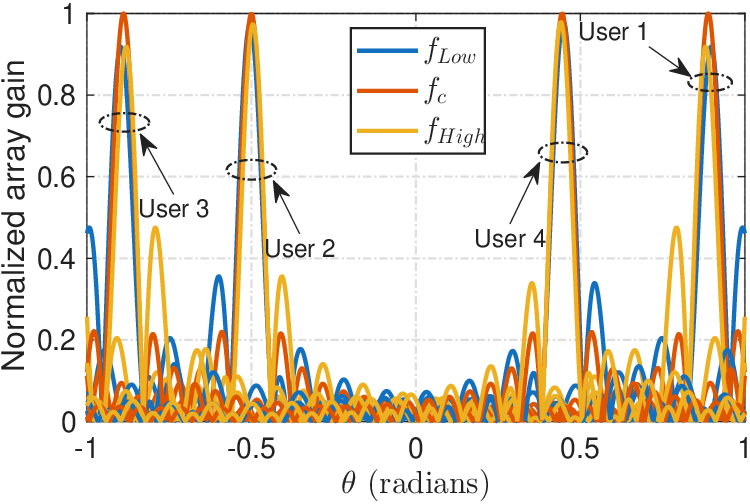}}
	\hfil
	\hspace{-20pt}\subfloat[]{\includegraphics[scale=0.38]{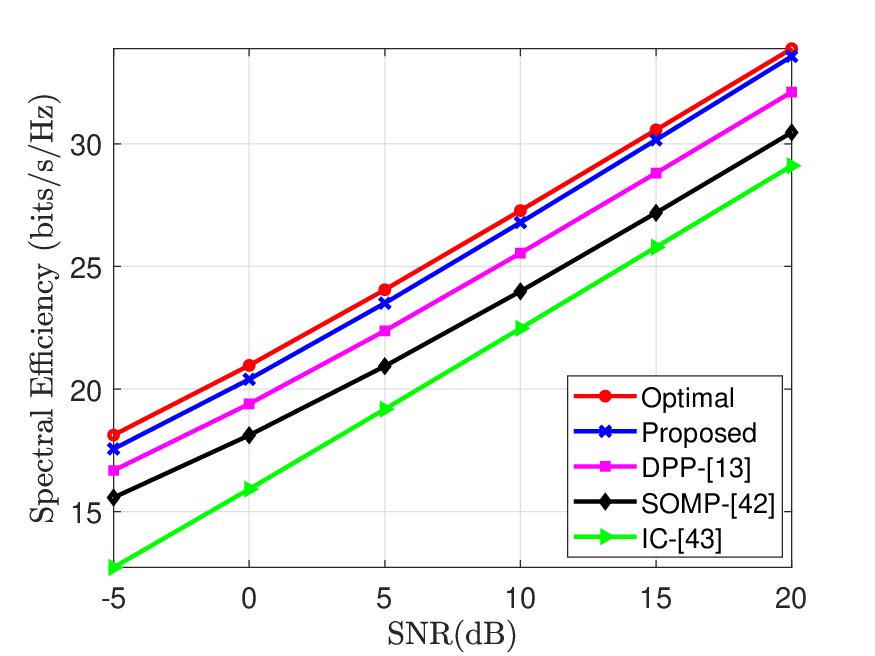}}
    \vspace{-2mm}
	\caption{$ \left(a\right) $ Normalized array gain vs physical direction corresponding to considered scenario  $ \left(b\right) $ Spectral efficiency in (bits/sec/Hz) vs SNR for the proposed and existing state-of-the-art approaches}\label{array_gain} \vspace{-1\baselineskip}
\end{figure*}
\vspace{-2mm}
\begin{align}
    \mathrm{GMM}(\mathsf{r}) = \frac{\mathsf{p_1}}{\sqrt{2 \pi }\nu_1}e^{-\frac{1}{2}\big(\frac{\mathsf{r}-\mathsf{r}_1}{\nu_1}\big)^2} + \frac{\mathsf{p_2}}{\sqrt{2 \pi }\nu_2}e^{-\frac{1}{2}\big(\frac{\mathsf{r}-\mathsf{r}_2}{\nu_2}\big)^2},
\end{align}
where $\mathsf{r}_1$ and $\mathsf{r}_2$ represent the actual physical approximation corresponding to the AoA/AoD. The quantity $\mathsf{p}_{(.)}$ represents the mixing weights uniformly distributed in the range $[0,1],$ and following $\mathsf{p}_2 = 1-\mathsf{p}_1,$ while $\nu_{(.)}$ represents the ray variance. The values of this are defined in Table-\ref{GMM_para}.
\begin{table}[hbt!]
    \centering
    \vspace{-2mm}
    \caption{GMM parameters for AoA/ AoD statistics \cite{priebe2011aoa}}
    \label{GMM_para}
    \resizebox{0.3\textwidth}{!}{
    \begin{tabular}{|c|c|c|} 
        \hline
        \textbf{Parameter} &  \textbf{AoA} $(\theta)$ & \textbf{AoD} $(\phi)$  \\ \hline
        $\mathsf{p_1}$ &  $0.724$ & $0.429$  \\ \hline
        $\mathsf{p_2}$ &  $2.198$ & $1.811$  \\ \hline
        $\nu_1$ &  $0.276$ & $0.571$  \\ \hline
        $\nu_2$ &  $7.297$ & $12.201$  \\ \hline
    \end{tabular}}\vspace{-0.8 \baselineskip}
\end{table}
The motivation for choosing the GMM over the Laplacian distribution \cite{garg2024angularly} lies in its ability to provide finer angular resolution and to capture a larger number of multipath components. Moreover, GMM can accommodate arbitrary reflection orders, which is particularly important in the THz band \cite{priebe2011aoa}. The quantity $\mathsf{p}_{(.)}$ represents the mixing weights uniformly distributed in the range $[0,1],$ and following $\mathsf{p}_2 = 1-\mathsf{p}_1,$ while $\nu_{(.)}$ represents the ray variance. The NMSE is given by $\mathrm{NMSE} = \frac{\sum_{k=0}^{K-1}\parallel \widehat{\mathbf{H}}[k]-\mathbf{H}[k]\parallel_\mathcal{F}^2}{\sum_{k=0}^{K-1}\parallel \mathbf{H}[k] \parallel_\mathcal{F}^2}$. Note that for all simulations, we employ the RRC-PSF-based dual-wideband channel model, unless stated otherwise. A trade-off analysis between the two channel formulations is also presented in Section-\ref{sim_channel}. Furthermore, to calculate the optimal digital precoders and combiners we follow the following procedure. Let the singular value decomposition (SVD) of the $u$-th user channel response $\mathbf{H}_u[k]$ can be obtained as
\begin{align}
    \mathbf{H}_u[k] = \mathbf{U}[k]\mathbf{\Sigma}[k]\mathbf{V}^H[k],
\end{align}
and the optimal transmit precoder $\mathbf{F}_u^{\mathrm{opt}}[k] \in \mathbb{C}^{N_u \times N_{s,u}}$ is comprised of the dominant $N_{s,u}$ columns of the right-singular matrix $\mathbf{V}[k]$ \cite{srivastava2021fast}. Let $\mathbf{H}_{\mathrm{eq},u}[k] = \mathbf{H}_u[k]\mathbf{F}_u^{\mathrm{opt}}[k] \in \mathbb{C}^{N_{\mathrm{BS}} \times N_{s,u}}$ represent the equivalent channel response. One can now formulate the optimal MMSE combiner $\mathbf{W}_{\mathrm{opt}}[k] \in \mathbb{C}^{N_{\mathrm{BS}} \times N_{s,u}}$
\begin{align}
    \mathbf{W}_{\mathrm{opt},u}[k] = \mathbf{H}_{\mathrm{eq},u}[k]\left(\mathbf{H}_{\mathrm{eq},u}^H[k]\mathbf{H}_{\mathrm{eq},u}[k] + \sigma^2 N_{s,u} \mathbf{I}_{N_{s,u}}\right)^{-1}. \notag
\end{align}
\subsection{Analysis of hybrid transceiver}
Fig. \ref{array_gain}(a) shows the normalized array gain for the first and third users as a function of their physical steering directions. It is evident that after applying the proposed TTD-based hybrid transceiver design, the array responses are frequency-aligned across all subcarriers. This demonstrates the ability of the TTD elements to effectively compensate for beam-squint by introducing frequency-dependent delays, thereby aligning the beams precisely with the user's physical directions at $\theta_1 \approx 27.85^\circ$ and $\theta_3 \approx 34.37^\circ$. As a result, consistent beam coverage is achieved over the entire bandwidth, validating the proposed design's efficiency in mitigating beam-squint effects in wideband THz systems.
\begin{figure*}
	\centering
	\hspace{-20pt}\subfloat[]{\includegraphics[scale=0.4]{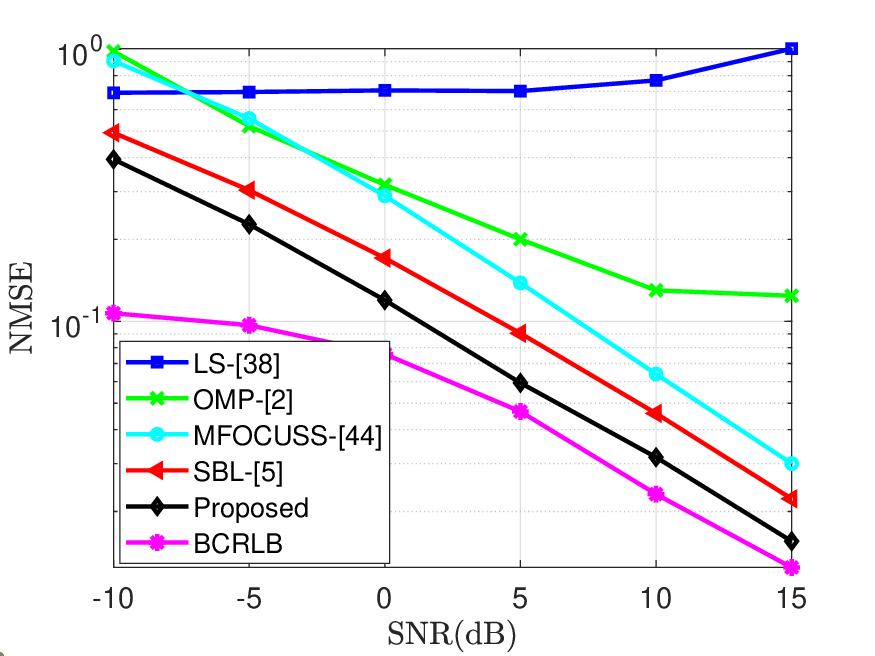}}
	\hfil
        \hspace{-20pt}\subfloat[]{\includegraphics[scale=0.4]{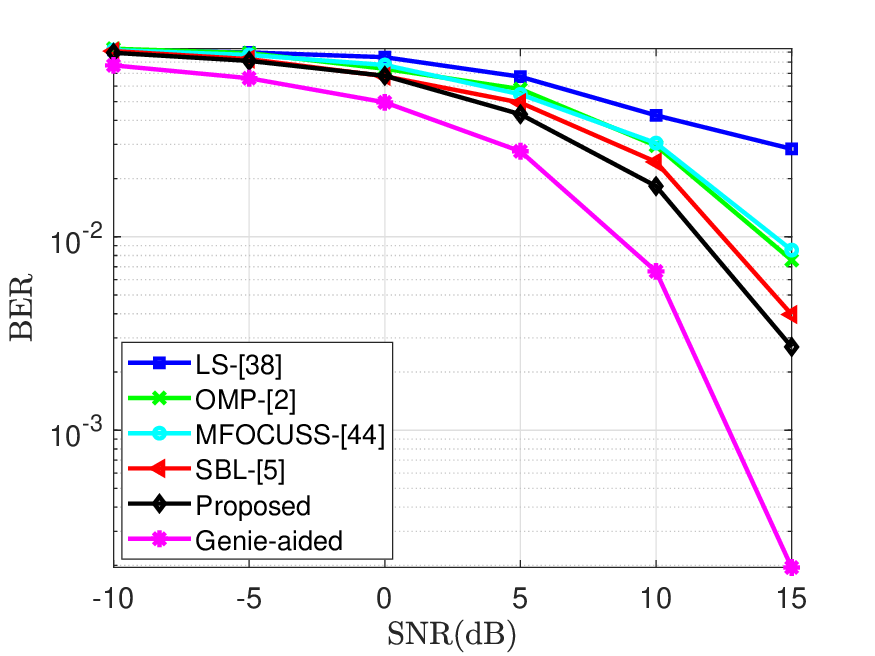}}
        \hfil
	\hspace{-20pt}\subfloat[]{\includegraphics[scale=0.4]{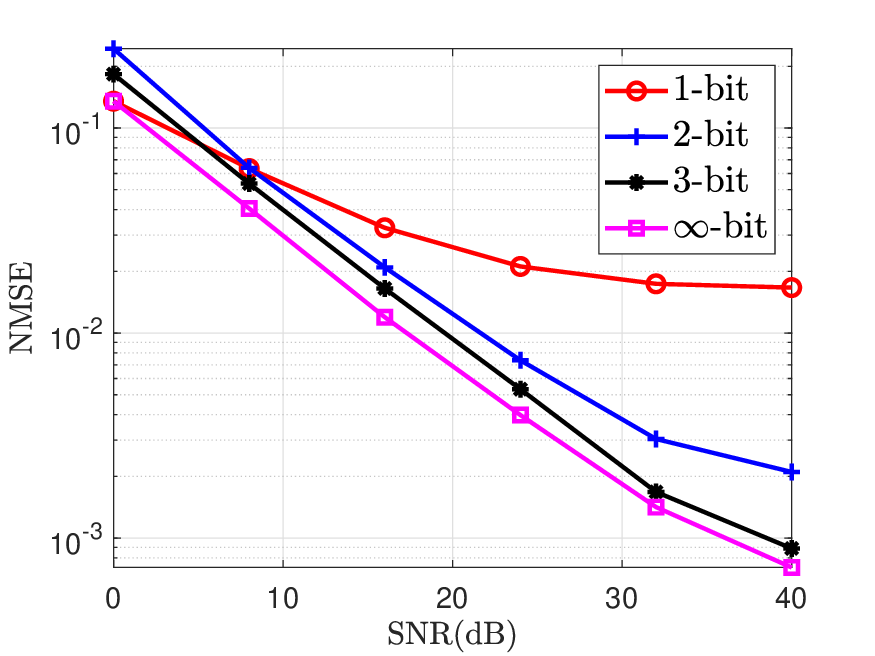}}
    \vspace{-2mm}
	\caption{$ \left(a\right) $ NMSE vs SNR comparison for the proposed and existing state-of-the-art approaches $ \left(b\right) $  BER vs SNR comparison for the proposed and existing state-of-the-art approaches $ \left(c\right) $ Effect of low-resolution ADCs on the proposed HBG-SR channel learning technique}\label{all_scheme}
\end{figure*}

Fig. \ref{array_gain}(b) depicts a substantial improvement in the achievable data rate when compared to existing state-of-the-art beamforming schemes, including DPP \cite{dai2022delay}, SOMP \cite{el2014spatially} and interference cancellation (IC) \cite{li2016robust} across varying SNR levels. The performance of SOMP and IC degrades significantly in the presence of beam-squint, since neither of these approaches incorporate a delay-compensation mechanism. Without delay elements their beamformers become frequency-dependent, leading to beam misalignment and reduced array gain across the bandwidth. The DPP method also suffers from performance limitations due to the absence of low-resolution modeling, making it less effective under coarse quantization, which is common in practical THz systems. By contrast, the proposed framework directly leverages the angular information embedded in the estimated beamspace, thereby eliminating the need for separate angle recovery and facilitating an integrated TTD-based hybrid transceiver design. This not only reduces computational complexity but also improves angular alignment across subcarriers, leading to significantly enhanced performance in wideband THz systems.
\vspace{-3mm}
\subsection{Analysis of channel estimation} \label{sim_channel}
This section presents the performance analysis of the proposed HBGR framework and compares it to the existing state-of-the-art techniques. Fig. \ref{all_scheme}(a) illustrates the NMSE performance as a function of SNR. The poor performance of the LS \cite{kay1993statistical} method is attributed to its inability to incorporate sparsity in the estimation process, which is a significant limitation compared to compressive sensing (CS) based techniques. Furthermore, the OMP algorithm \cite{venugopal2017channel} suffers from both structural and convergence errors. Specifically, a low stopping threshold leads to many iterations and significant nonzero components, causing structural errors. Conversely, a high stopping threshold reduces iterations, but fails to capture all dominant channel components, potentially resulting in convergence errors. Moreover, the MMV Focal Underdetermined System Solver (MFOCUSS) algorithm \cite{wipf2007empirical} often converges to suboptimal local minima, resulting in convergence errors. The Bayesian framework leverages the EM algorithm for hyperparameter learning, eliminating the need for manual tuning or regularization and turns out to be robust the choice of the dictionary matrix. The conventional Bayesian learning technique \cite{garg2024angularly} estimates the channel support independently for each subcarrier. In wideband systems affected by beam-squint, the effective steering vectors change with frequency, causing the support to vary across subcarriers. Since the exact supports differ, these variations cause SBL to perform poorly in this scenario. By contrast, the proposed HBG-SR approach enforces joint sparsity by associating a shared group of hyperparameters with each angle bin across all subcarriers. This leverages the underlying group-sparse structure of the channel, facilitating more consistent support detection and producing significantly lower NMSE, as observed in the simulation results. Moreover, the performance of the proposed method closely approaches the BCRLB, which is particularly noteworthy given that the BCRLB is derived under the idealized assumption of perfect AoA/AoD knowledge. Note that the BCRLB characterizes a fundamental lower bound on the achievable MSE under the assumed probabilistic channel model and prior, but does not by itself imply that the bound is always tight or exactly attainable by practical estimators. Prior studies \cite{niazadeh2011achievability}, \cite{ben2010cramer} have shown that Cram\'er Rao type bounds are asymptotically tight in regimes, where the effective support is correctly identified and the SNR is sufficiently high. In particular, for sparse parameters associated with maximal support, the CRB coincides with the oracle bound and it is achievable by the maximum likelihood estimator in the high-SNR limit, as also reflected by the simulation results of Fig. \ref{all_scheme}(a). However, in the THz hybrid MIMO setting considered, finite-SNR operation, model mismatch, and hardware impairments generally prevent exact achievability of the BCRLB. Therefore, the bound should be interpreted as a benchmark for best-case performance.
\begin{figure*}
	\centering
\subfloat[]{\includegraphics[scale=0.32]{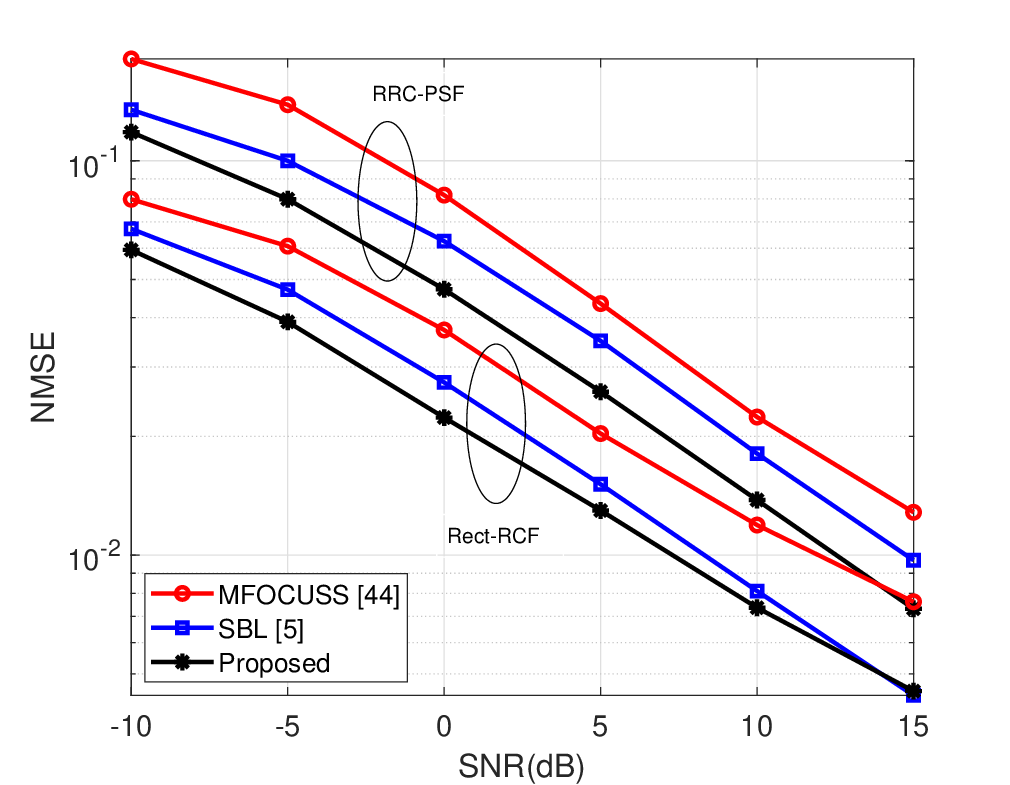}}
	\hfil
	\hspace{-20pt}\subfloat[]{\includegraphics[scale=0.38]{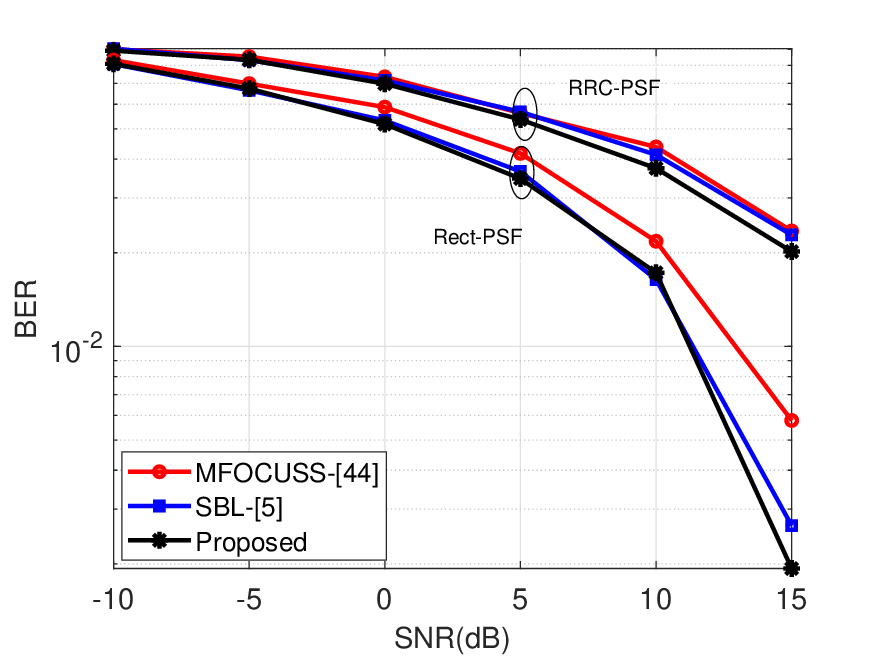}}
        \hfil
	\hspace{-20pt}\subfloat[]{\includegraphics[scale=0.38]{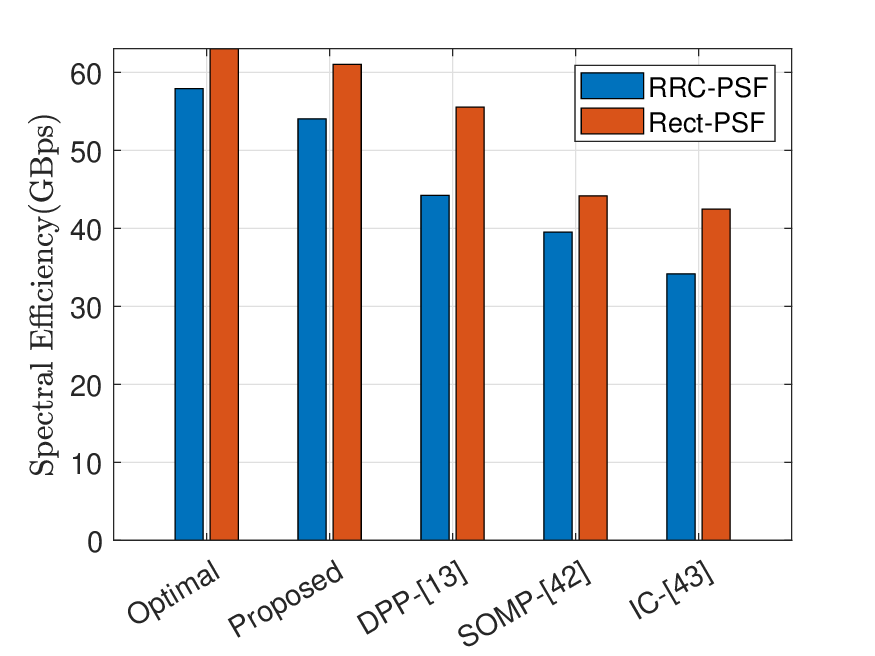}}
    \vspace{-2mm}
	\caption{Performance comparison for the proposed HBG-SR technique with RRC-PSF and Rect-PSF based dual-wideband channel formulations $ \left(a\right) $ NMSE vs SNR  $ \left(b\right) $ BER vs SNR $ \left(c\right) $ Spectral efficiency with different transceiver schemes.}\label{pcf_rcf} \vspace{-1.2 \baselineskip}
\end{figure*}

Fig \ref{all_scheme}(b) depicts the accuracy of data detection achieved by employing the Minimum Mean Squared Error (MMSE) receiver with the proposed HBG-SR technique, along with other conventional sparse estimation based techniques and the Genie-aided detector as benchmarks. As illustrated, the resultant BER decreases consistently upon increasing the SNR, which closely aligns with the NMSE results presented previously, underscoring the high quality of the CSI estimated. Additionally, the BER of the proposed channel learning approach closely matches the hypothetical Genie-aided detector, highlighting its efficacy and robustness in realistic scenarios. Fig \ref{all_scheme}(c) evaluates the effect of employing low-resolution ADCs on the CSI estimation accuracy. It can be observed that even with $3$-bit ADC resolution, the proposed algorithm attains a performance close to an ideal $\infty$-resolution transceiver scenario. This result holds significant practical value, demonstrating the robustness of our channel learning technique in scenarios, where low-resolution ADCs are essential for managing high bandwidth with limited power budgets.
\subsection{Performance comparison for RRC-PSF and Rect-PSF based dual-wideband formulations}
To establish a baseline for comparison and to highlight the
performance trade-offs introduced by practical pulse shaping, we also consider the Rect-PSF based dual-wideband channel
model, as adopted in \cite{dovelos2021channel}, \cite{chou2023compressed}. Fig. \ref{pcf_rcf}(a) compares the performance of the HBG-SR based channel learning approach by considering both the RRC-PSF and Rect-PSF based dual-wideband channels. The Rect-PSF based dual-wideband channel suffers from spectral leakage, which leads to a higher level of interference between the adjacent channels, which may erode the overall performance of the communication system. On the other hand, the RRC-PSF based dual wideband channel introduces the problem of signal strength reduction at the edges of symbol duration. This attenuation may adversely affect signal recovery at the receiver end, as the diminished signal may fall below the receiver’s detection threshold. Thus, there is a clear trade-off between both the models and the consideration of the filter depends upon the specific application. Fig. \ref{pcf_rcf}(b) compares the BER performance of the proposed HBG-SR-based channel learning framework over dual-wideband THz channels modeled using the RRC-PSF and Rect-PSF filters. In both cases, it is evident that the HBG-SR-based approach consistently outperforms the conventional SBL and MFOCUSS techniques across the entire SNR range. Notably, the improvement is more prominent in the Rect-PSF-based dual-wideband channel due to its flat response for the entire symbol duration, which exacerbates beam-squint and frequency-dependent distortions. The robustness of the proposed method under both pulse-shaping filters highlights its practical applicability and makes it a strong candidate.

Fig. \ref{pcf_rcf}(c) compares the achievable spectral efficiency (in Gbps) of the proposed TTD-based hybrid beamforming scheme to several baseline methods, including the SOMP and IC technique, under both the RRC-PSF and Rect-PSF based dual-wideband channel models. The proposed approach consistently achieves higher throughput for both pulse-shaping filters and closely approaches the performance of the fully digital optimal benchmark. This gain is attributed to the TTD elements, which effectively compensate for frequency-dependent phase misalignments and mitigates the beam-squint effect across wideband subcarriers. Notably, all simulations are conducted under practical hardware constraints with $3$-bit ADC resolution, highlighting the robustness and efficiency of the proposed design in low-resolution scenarios. These results validate the framework’s capability to jointly address wideband beamforming and channel estimation challenges in THz systems with dual-wideband impairments. Potential future research includes the incorporation of blockage effects and molecular absorption noise to further investigate the performance of the proposed framework under realistic THz propagation conditions.
\vspace{-2mm}
\section{Conclusion} \label{conclusions}
\vspace{-1mm}
A novel HBG-SR-based channel estimation framework for wideband THz MIMO systems was proposed. To accurately capture the wideband propagation characteristics, we formulated a dual-wideband channel model incorporating RRC-PSF. The Bussgang decomposition was employed to linearize the nonlinear signal model arising from low-resolution ADCs, facilitating tractable inference under coarse quantization. Leveraging the estimated beamspace information, a hybrid transceiver design based on TTD was presented to mitigate the beam-squint effect and enable frequency-flat beam alignment across subcarriers. Simulation results demonstrated that the proposed approach consistently outperforms existing techniques in terms of channel estimation accuracy, BER and achievable data rate, while remaining efficient under low-resolution hardware constraints. Overall, the joint design offers a scalable, robust, and hardware-aware solution for next-generation wideband THz communication systems.
\vspace{-2mm}
\appendices
\section{Calculation of Noise Covariance} \label{noise_covariance_derivation}
\vspace{-1mm}
Revisiting Eq. \eqref{received_signal}, the received signal before passing through a low-resolution ADC can be expressed as
\vspace{-2mm}
\begin{align}
    \acute{\mathbf{y}}_m(q)=\mathbf{W}_{\mathrm{RF},m}^H \sum_{u=1}^U\mathbf{H}_{q,u} \otimes_k \tilde{\mathbf{g}}_{m,u}^{(q)}+\mathbf{W}_{\mathrm{RF},m}^H{\mathbf{v}}_m(q). \label{before_ADC}
\end{align}
where $\mathbb{E}(\tilde{\mathbf{g}}_{m,u}(p)\tilde{\mathbf{g}}_{m,u}^H(p))= \mathbf{T}_{\tilde{g}\tilde{g},m} = \sigma_d^2\mathbf{F}_{\mathrm{RF},m,u}\mathbf{F}_{\mathrm{BB},m,u}\mathbf{F}_{\mathrm{BB},m,u}^H\mathbf{F}_{\mathrm{RF},m,u}^H$. Expanding the circular convolution, as defined in Section-\ref{nota}, the above equation can be rewritten as
\vspace{-2mm}
\begin{align}
    &\acute{\mathbf{y}}_m(q)  = \mathbf{W}_{\mathrm{RF},m}^H\sum_{u=1}^U\big[\mathbf{H}_u(0)\tilde{\mathbf{g}}_{m,u}(q)+\mathbf{H}_u(1)\tilde{\mathbf{g}}_{m,u}(q-1)+ \notag \\ & \cdots+\mathbf{H}_u(K-1)\tilde{\mathbf{g}}_{m,u}(q-K+1)\big]+\mathbf{W}_{\mathrm{RF},m}^H{\mathbf{v}}_m(q). \label{rec_vc}
\end{align}
Let the received signal's covariance matrix $\tilde{\mathbf{J}}_m \in \mathbb{C}^{N_{RF}^B \times N_{RF}^B}$ be defined as $\tilde{\mathbf{J}}_m = \mathbb{E}\{ \acute{\mathbf{y}}_m(q)\acute{\mathbf{y}}_m^H(q)\}$. After substituting Eq. \eqref{rec_vc} in $\tilde{\mathbf{J}}_m$, one obtains
\vspace{-1mm}
\begin{align}
    &\tilde{\mathbf{J}}_m = \mathbf{W}_{\mathrm{RF},m}^H \sum_{u=1}^H \big[ \mathbf{H}_u(0)\mathbf{T}_{\tilde{g}\tilde{g},m}\mathbf{H}_u^H(0) + \cdots + \notag \\ &\mathbf{H}_u(K-1)\mathbf{T}_{\tilde{g}\tilde{g},m}\mathbf{H}_u^H(K-1) \big] \mathbf{W}_{\mathrm{RF},m}^H + \sigma_n^2\mathbf{W}_{\mathrm{RF},m}^H\mathbf{W}_{\mathrm{RF},m}, \notag
\end{align}
which can be compactly written as $\tilde{\mathbf{J}}_m = \mathbf{W}_{\mathrm{RF},m}^H\mathbf{L}_m\mathbf{W}_{\mathrm{RF},m}+\sigma_n^2\mathbf{W}_{\mathrm{RF},m}^H\mathbf{W}_{\mathrm{RF},m},$ where we have $\mathbf{L}_m=\sum_{u=1}^U\sum_{n=0}^{K-1}\mathbf{H}_u(n)\mathbf{T}_{\tilde{g}\tilde{g},m}\mathbf{H}_u^H(n)$. Therefore, the noise covariance matrix $\mathbf{J}_m \in \mathbb{C}^{N_{RF}^B \times N_{RF}^B}$ after passing through a low-resolution ADC \cite{jacobsson2017throughput} can be expressed as $\mathbf{J}_m=\varkappa(1-\varkappa)\mathrm{diag}(\tilde{\mathbf{J}}_m)$.
\section{Calculation of Maximum Array Gain} \label{MAG}
Substituting Eq. \eqref{array_res} into \eqref{nor_arr_gain}, one obtains
\begin{align}
    \Upsilon = \frac{1}{N}\left|\begin{bmatrix}
        1 \\ e^{-j \pi \varrho_k \sin \theta_k} \\ \vdots \\ e^{-j \pi (N-1) \varrho_k \sin \theta_k}
    \end{bmatrix}^H \begin{bmatrix}
        1 \\ e^{-j \pi \sin \theta} \\ \vdots \\ e^{-j \pi \left(N-1 \right) \sin \theta}
    \end{bmatrix} \right|.
\end{align}
Further simplifying the above equation and assuming $\theta$ is small, we obtain
\begin{align}
    \Upsilon = \frac{1}{N}\Big|\sum_{n=0}^{N-1}e^{jn\pi(\varrho_k \theta_k - \theta)} \Big|.
\end{align}
Therefore, in order to achieve maximum array gain, $\theta_k = \frac{\theta}{\varrho_k}$.
\section{Calculation of optimal angle} \label{opti_cal}
Revisiting Eq. \eqref{nor_arr_gain} and substituting the array response vector as given in Eq. \eqref{arry_res_TTD} into the normalized array gain expression of Eq. \eqref{nor_arr_gain}, one obtains
\vspace{-2mm}
\begin{align}
    \Upsilon_k = \big|(\tilde{\mathbf{a}}(\theta,f_c))^H \tilde{\mathbf{a}}_k(\theta_k,f_k)\big|,
\end{align}
\begin{figure*}
    \begin{align}
    \Upsilon_k  = \bigg|\frac{1}{N}  \Big\{\big(1+ & e^{j \pi \theta}e^{-j\pi\varrho_k\theta_k}+  \cdots+e^{j \pi (P-1)\theta}e^{-j\pi(P-1)\varrho_k\theta_k}\big) +  e^{-j\pi v_k} \big(1+e^{j \pi P \theta}e^{-j\pi P \varrho_k\theta_k}+ \cdots+e^{j \pi (2P-1)\theta}e^{-j\pi(2P-1)\varrho_k\theta_k} \big) \notag \\ &  + \cdots + e^{-j\pi (S-1) v_k} \big(1+e^{j \pi (S-1)P \theta}e^{-j\pi (S-1)P v_k\theta_k}+ \cdots +  e^{j \pi (SP-1)\theta}e^{-j\pi(SP-1)\varrho_k\theta_k} \big) \Big\} \bigg|, \label{normalize}
\end{align}
\vspace{-7mm}
\begin{align}
    \Upsilon_k  = \frac{1}{N} \bigg| \sum_{m=1}^S \sum_{p=1}^P e^{j \pi [(s-1)P+(s-1)]\theta} e^{-j \pi \varrho_k[(s-1)P+(p-1)]\theta_k} e^{-j\pi(s-1)v_k}\bigg|, \label{normalize_angle}
\end{align}
\hrulefill
\end{figure*}
which can be further simplified to Eq. \eqref{normalize}-\eqref{opt_ang}.
\vspace{-1mm}
\begin{align}
    \Upsilon_k  = \frac{1}{N} \bigg| \sum_{s=1}^S e^{j\pi(s-1)\left\{P\theta - \varrho_k P \theta_k -v_k \right\}} \sum_{p=1}^P e^{-j\pi(p-1)\left\{-\theta+\varrho_k\theta_k \right\}} \bigg|. \label{opt_ang}
\end{align}
Therefore, the normalized array gain can be represented in the form of the Dirichlet sinc function, as
\begin{align}
    \Upsilon_k = \frac{1}{N} \big|\Xi_S\big(P\theta - \varrho_k P \theta_k -v_k \big) \Xi_P \big(-\theta+\varrho_k\theta_k \big)\big|. \label{dir_array}
\end{align}
Thus, the optimal angle at which the array gain is maximum can be derived by substituting $P\theta - \varrho_k P \theta_k -v_k = 0$ into Eq. \eqref{dir_array} which finally yields Eq. \eqref{optimal_angle}.
\balance
\bibliographystyle{IEEEtran}
\bibliography{References}
\end{document}